\documentclass[pdflatex,sn-mathphys,referee]{sn-jnl}% Math and Physical Sciences Reference Style
\jyear{2022}%

\theoremstyle{thmstyleone}%
\theoremstyle{thmstyletwo}%

\theoremstyle{thmstylethree}%

\begin{document}

\title[Inferring surface elastic moduli in thin walls of single cells]{Inferring relative surface elastic moduli in thin-wall models of single cells}


\author[1]{\fnm{Yaqi} \sur{Deng}}\email{yaqi.deng21@imperial.ac.uk}
\equalcont{These authors contributed equally to this work.}

\author[2]{\fnm{Chaozhen} \sur{Wei}}\email{cwei4@wpi.edu}
\equalcont{These authors contributed equally to this work.}
\author[3]{\fnm{Rholee} \sur{Xu}}\email{rxu3@wpi.edu}
%%\equalcont{These authors contributed equally to this work.}
%
\author[3,4]{\fnm{Luis} \sur{Vidali}}\email{lvidali@wpi.edu}
\author*[2,3]{\fnm{Min} \sur{Wu}}\email{englier@gmail.com}

\affil[1]{\orgdiv{Life Sciences}, \orgname{Imperial College London}, \orgaddress{\city{London}, \postcode{SW7 2AZ}, \country{UK}}}
\affil[2]{\orgdiv{Mathematical Sciences}, \orgname{Worcester Polytechnic Institute}, \orgaddress{\city{Worcester}, \postcode{01609}, \state{MA}, \country{USA}}}

\affil[3]{\orgdiv{Bioinformatics and Computational Biology}, \orgname{Worcester Polytechnic Institute}, \orgaddress{\city{Worcester}, \postcode{01609}, \state{MA}, \country{USA}}}

\affil[4]{\orgdiv{Biology and Biotechnology}, \orgname{Worcester Polytechnic Institute}, \orgaddress{\city{Worcester}, \postcode{01609}, \state{MA}, \country{USA}}}
%
%\affil[3]{\orgdiv{Department}, \orgname{Organization}, \orgaddress{\street{Street}, \city{City}, \postcode{610101}, \state{State}, \country{Country}}}


\abstract{
There is a growing interest in measuring the cell wall mechanical property at different locations in single walled cells. We present an inference scheme that maps relative surface elastic modulus distributions along the cell wall based on tracking the location of material marker points along the turgid and relaxed cell wall outline. A primary scheme is developed to provide a step-function inference of surface elastic moduli by computing the tensions and elastic stretches between material marker points. We perform analysis to investigate the stability of the primary scheme against perturbations on the marker-point locations, which may occur due to image acquisition and processing from experiments. The perturbation analysis shows that the primary scheme is more stable to noise when the spacing between the marker points is coarser, and has been confirmed by the numerical experiments where we apply the primary scheme to synthetic cell outlines from simulations of hyper-elastic membrane deformation with random noise on the marker-point locations. To improve the spatial resolution of elastic modulus distribution of the primary scheme with noise, we propose two optimization schemes that convert the step-function inferences of elastic moduli into smooth-curve inferences. The first optimization scheme infers a canonical elastic modulus distribution based on marker-point locations from multiple cell samples of the same cell type. The second optimization scheme is a simplified cost-effective version that infers the elastic moduli based on marker-point locations from a single cell. The numerical experiments show that the first optimization scheme significantly improves the inference precision for the underlying canonical elastic modulus distributions and can even capture some degree of nonlinearity when the underlying elastic modulus gradients are nonlinear. The second cost-effective scheme is capable of predicting the trend of the elastic modulus gradients consistently.  
}

%\keywords{keyword1, Keyword2, Keyword3, Keyword4}



\maketitle

\section{Introduction}\label{sec:Intro}

In plants, fungi, and bacteria, cell walls play an essential role in cell development and adaptation---the cell wall functions as a boundary and a supporting structure that opposes the internal turgor pressure resulting from osmosis. Thus, controlling the mechanical properties of the cell wall is of critical importance for walled cell development and survival. In growing filaments such as root hairs, pollen tubes, and fungus hyphae, the cell wall mechanics affects their shape and extension rate \cite{dumais2004mechanics,dumais2006anisotropic,campas2009shape,campas2012strategies,rojas2011chemically,goriely2003self,fayant2010finite}. Hence, measuring the cell wall mechanical properties within one cell or among different cells within one species can improve our understanding of how cells develop and adapt to external environmental cues. 

As for single plant cells, the bulk modulus of cells $\epsilon=P/\frac{dV}{V}$ has been measured in various systems \cite{1987pressure,buchner1981turgor,murphy1995new} by manipulating the pressure $P$ inside the cell while tracking the fraction change in the cell volume $\frac{dV}{V}$. Recently in fission yeast \cite{abenza2015wall,atilgan2015morphogenesis,davi2019systematic}, the cell-wall elastic property is distinguished from the cell bulk, where the cell wall is assumed to be elastic in the lateral directions along the wall surface. The surface elastic constants such as Young's modulus and Poisson's ratio $\nu$ have been inferred by measuring the elastic strains along the wall surface and applying Young-Laplace's Law, which connects the local surface curvature with the lateral tensions and the turgor pressure. In \cite{davi2019systematic}, it is reported that the Young's moduli at specific landmarks on the fission yeast wall (e.g., the two ends, sides, and scars) are different.  In addition to the results in fission yeasts, microindentation experiments \cite{chebli2012cell} and numerical simulations \cite{bolduc2006finite,fayant2010finite} suggest that the stiffness of the cell wall varies from the tip end to the flatter regions in pollen tubes. Thus, we should further map the spatial distribution of the elastic constants along the cell wall surface.

Previously in \cite{chelladurai2020inferring}, we have developed an inference scheme that maps relative lateral tensions along the cell wall based on the cell outline coordinates. This approach is similar to other tension-inference procedures \cite{hejnowicz1977tip,dumais2004mechanics,rojas2011chemically}, except that we have analyzed how this inference scheme responds to noise quantitatively. We found that a coarser spatial discretization of the cell outline is required to generate reliable results as noise amplitude on the cell-outline registration increases. To effectively restore the spatial resolution of the primary scheme that is more robust with low spatial resolution, we further developed an optimization method by converting the discrete tension distributions to smooth polynomials. 

In this paper, by expanding the previous work \cite{chelladurai2020inferring}, we develop an inference scheme that maps relative surface elastic moduli along the cell wall based on the cell outline coordinates. The full inference approach presented in this study includes 1) the primary scheme that maps the elastic moduli on the discretized cell outlines as step functions and 2) the optimization schemes that give the approximated smooth representations of the primary inference. Similar to \cite{chelladurai2020inferring}, we have studied the stability of the primary scheme against small noise analytically. By applying the primary scheme to synthetic cell outlines from a computational model of hyper-elastic membrane deformation with random noise, we have found agreement between the error bounds predicted by the analysis and the error computed from numerical experiments. In the end, we present two optimization methods where the first is suitable to infer the canonical elastic modulus distributions integrating multiple cell outline data and the second is a simplified cost-effective version which infers the elastic moduli within one cell.

The paper is organized as follows. In Sec.~2, we provide the background of the geometry and mechanics in axially-symmetric cell walls, propose a hyper-elastic constitutive law to describe the elastic property of the cell wall due to stretching of the wall surface, and discuss its connection with linear elastic laws used in previous works \cite{abenza2015wall,atilgan2015morphogenesis}. In Sec.~3, we formulate the basic inference scheme that computes the surface elastic moduli as step-function distributions based on the cell outline coordinates in both the turgid and relaxed configurations, and perform sensitivity analysis to this scheme. In particular, we provide formulae of how small cell outline perturbation propagates to the tensions and elastic strains, and how the errors in the tension and strain measurements affect the inferred surface bulk and shear moduli quantitatively. In Sec.~4, we apply the scheme to infer the surface elastic modulus distributions from cell outlines generated by a computational model. We show the relative error between the inferred elastic moduli and the prescribed elastic moduli agrees within the error bound distribution predicted by the sensitivity analysis.  In Sec.~5, we formulate an auxiliary scheme that converts the primary step-function elastic modulus measurements from multiple numerical experiments to one pair of polynomial representations of bulk and shear modulus distributions. In Sec.~6, we formulate a simplified scheme that can be applied to the cell outline data of a single cell. In Sec.~7, we discuss our results with their connections to the molecular composition and structure of the cell wall, the limitation of the image-based elastic-moduli inference schemes, and their future applications. 

\section{Cell wall geometry and mechanics}\label{sec:mechanics}
\begin{figure}[t]
	\centering
	\includegraphics[width=1\linewidth]{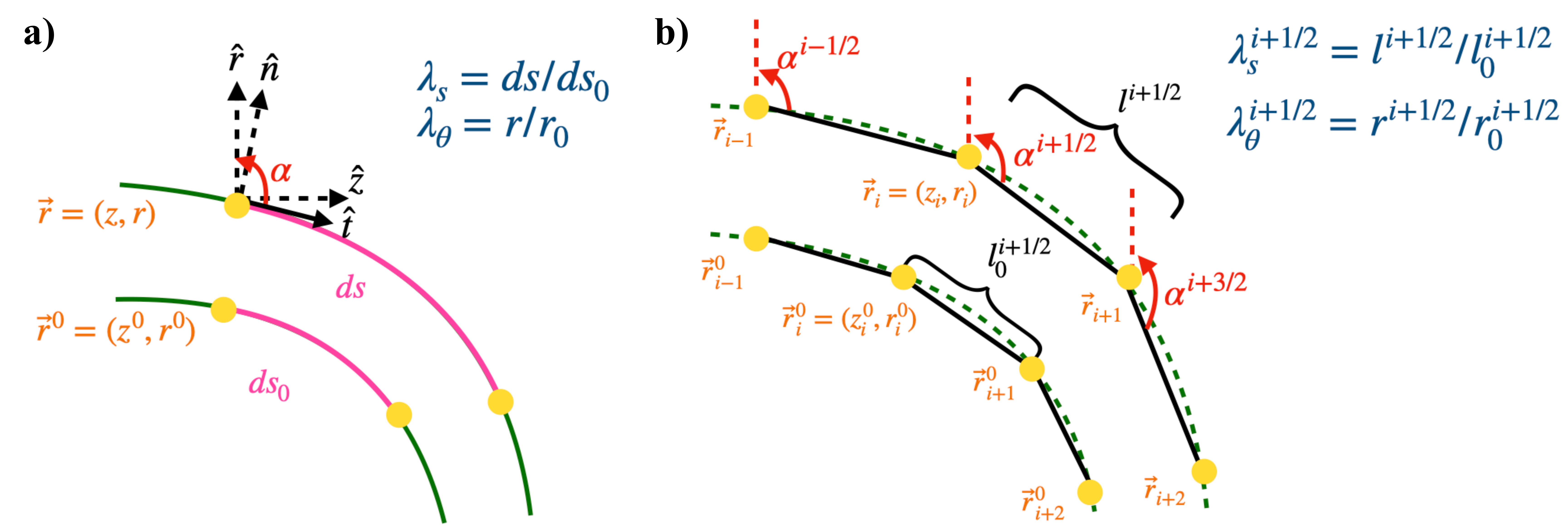}
	\caption{\textbf{Illustration of cell wall representation}. \textbf{a)} The cell wall is represented by the curve $\vec{r}(s)=(z(s), r(s))$ and $\vec{r}^0(s)=(z(s^0), r(s^0))$ in the turgid and relaxed configurations respectively, where $s$ and $s^0$ represent the arc length from a reference point. $\lambda_s$ and $\lambda_\theta$ are the elastic stretches in the meridional and circumferential directions. \textbf{b)} The discretization of cell outline by marker points $\vec{r}_i=(z_i,r_i)$ and $\vec{r}_i^0=(z^0_i,r^0_i)$, in the turgid and relaxed configurations. }
	\label{wallMechanics}
\end{figure}

Since tip cells of plants have a nearly symmetrical and smooth apex, in our model the cell wall is described as a thin shell structure that is a surface of revolution about the axis of symmetry. Assumed as a compressible hyper-elastic material, the cell wall undergoes elastic deformation from the initial relaxed configuration, $\vec{r}^0(s^0)=(z^0(s^0), r^0(s^0))$, to the turgid configuration, $\vec{r}(s)=(z(s), r(s))$, due to the turgor pressure $P$, where $z$ ($z^0$) is the coordinate along the axis of symmetry, $r$ ($r^0$) is the local radius from the axis of symmetry, and $s$ ($s^0$) is the arc length from a reference point at the rear boundary in the turgid (relaxed) configuration (Fig.~\ref{wallMechanics}a). By the force balance between the turgor pressure $P$ and the tensions $\sigma_s$ and $\sigma_\theta$ along the meridional and circumferential directions \cite{chelladurai2020inferring}, respectively, we can derive the widely used Young-Laplace Law \cite{goriely2003self,dumais2006anisotropic,campas2009shape}: $\kappa_s\sigma_s+\kappa_\theta\sigma_\theta=P$ and $\kappa_\theta\sigma_s=P/2$, where $\kappa_s$ and $\kappa_\theta$ are the curvatures along the meridional and cirfumferential directions, respectively. The Young-Laplace Law gives a way to retrieve the tensions based on the turgor pressure and the geometry of the tip-cell
\begin{align}\label{eq:stress_curvature1}
&\sigma_s=P/(2\kappa_\theta),\\ \label{eq:stress_curvature2}
&\sigma_\theta=P/(2\kappa_\theta)\times(2-\kappa_s/\kappa_\theta),
\end{align}
where the curvatures can be easily calculated by $\kappa_s=d\alpha/ds$ and $\kappa_\theta=\sin\alpha/r$, where $\alpha$ is defined as the angle from the local tangent vector $\hat{t}$, in the direction of increasing $s$, to the radial direction $\hat{r}$ (Fig.~\ref{wallMechanics}a). From now on, we assume a constant turgor pressure $P\neq0$ along the cell wall.

In addition to the relation with local curvatures, the tensions are related with the elastic stretches by the constitutive law. In this paper, we assume a compressible neo-Hookean constitutive law \cite{chelladurai2020inferring} (see Sec.~\ref{sec:discussion} for discussion of constitutive laws)
\begin{align}\label{eq:continuous_tension1}
&\sigma_s=\frac{1}{2}\mu_h(\frac{1}{\lambda_\theta^2}-\frac{1}{\lambda_s^2})+K_h(\lambda_s\lambda_\theta-1),\\ \label{eq:continuous_tension2}
&\sigma_\theta=\frac{1}{2}\mu_h(\frac{1}{\lambda_s^2}-\frac{1}{\lambda_\theta^2})+K_h(\lambda_s\lambda_\theta-1),
\end{align}
where $\lambda_s$ and $\lambda_\theta$ are respectively the elastic stretches in the meridional and circumferential directions, $\mu_h=h \times \mu$ and $K_h=h \times K$ are respectively the rescaled shear and bulk elastic moduli by the thickness of the cell wall $h$ in the thin-shell approximation, and hence the tensions $\sigma_s$ and $\sigma_\theta$ are the in-plane stresses multiplied by the thickness of cell wall. This constitutive law based on nonlinear elasticity can describe the mechanical behavior of a wide range of transversely isotropic biomaterials, which is the case of many cells with tip growth (see Discussion for more details). In particular, this nonlinear constitutive law reduces to the linear elastic law used in the previous work for fission yeast \cite{abenza2015wall,atilgan2015morphogenesis} when the elastic strains $\epsilon_s=\lambda_s-1$ and $\epsilon_\theta=\lambda_\theta-1$ are small, and the shear and bulk modulus can be related to the rescaled Young's modulus $E_h=h\times E$ and Poisson ratio $\nu$ via $\mu_h=\frac{E_h}{2(1+\nu)}$ and $K_h=\frac{E_h}{2(1-\nu)}$.

\begin{figure}[t]
	\centering
	\includegraphics[width=1\linewidth]{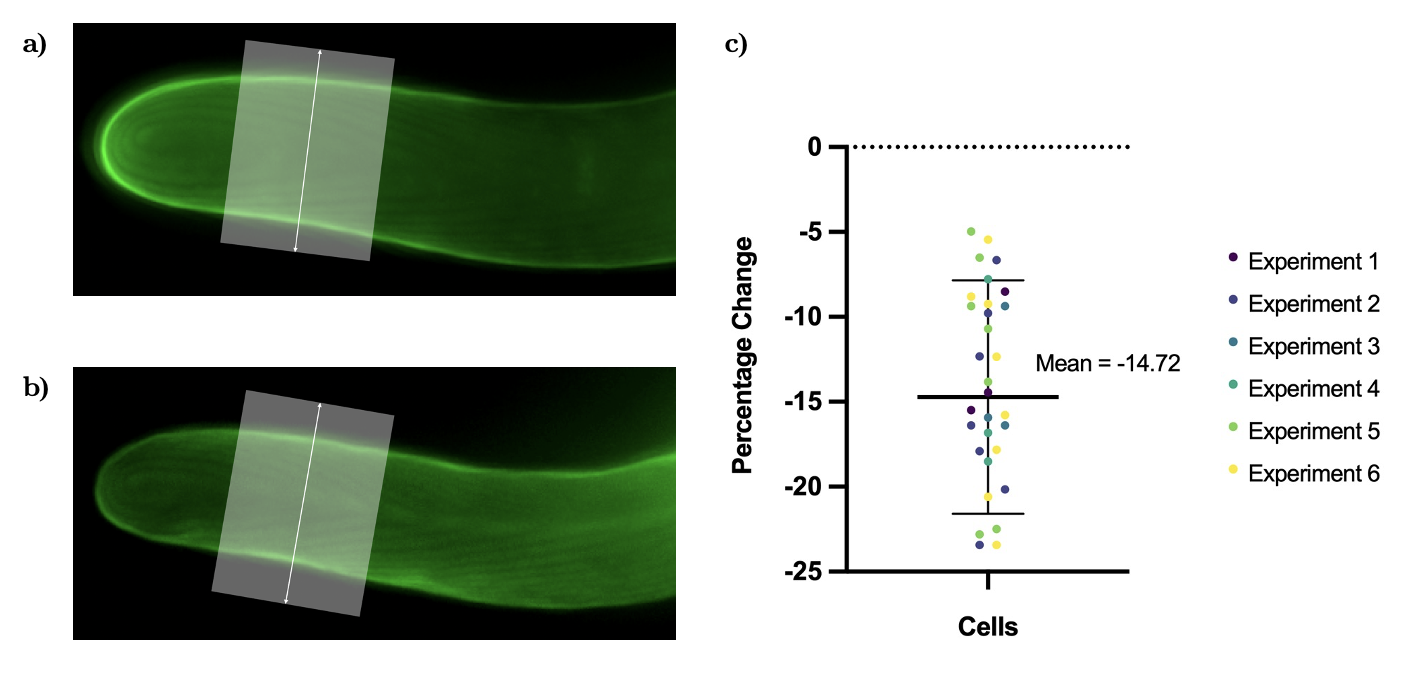}
	\caption{\textbf{\textit{Physcomitrium patens} caulonema cell with and without turgor pressure.} \textbf{a)} Caulonema tip cell under normal conditions with cell wall outlined with Direct Yellow 96 dye. White box depicts where the brightness profile was plotted to calculate the width of the cell. \textbf{b)} Same cell after a treatment of 15$\%$ Mannitol in liquid media. \textbf{c)} Summary of percent width change in cells before and after plasmolysis, where the percent change is calculated with (Width after $–$ Width before)/(Width after)$\times$100. Outlier of -37$\%$ is not shown in the graph, but is included to calculate the mean. The total number of cells measured was 31, spread across 6 different experiments. Error bars indicate standard deviation. See Appendix~\ref{sec:S0} for experimental protocols and details of cell width measurements.  }
	\label{fig:moss_image}
\end{figure}

By the constitutive law (\ref{eq:continuous_tension1}) and (\ref{eq:continuous_tension2}), the elastic moduli of the cell wall are related to the tensions and elastic stretches by
\begin{align}\label{eq:continuous_K}
&K_h=\frac{\sigma_s+\sigma_\theta}{2(\lambda_s\lambda_\theta-1)},\\
\label{eq:continuous_mu}
&\mu_h=\frac{\sigma_s-\sigma_\theta}{1/\lambda_\theta^2-1/\lambda_s^2},
\end{align}
where the tensions $\sigma_s$ and $\sigma_\theta$ are retrieved from the turgor pressure and the turgid cell-wall shape by Eqs.~(\ref{eq:stress_curvature1}) and (\ref{eq:stress_curvature2}), and the elastic stretches $\lambda_s=ds/ds_0$ and $\lambda_\theta=r/r_0$ are computed by comparing the relaxed and turgid configruations of cell wall. We have measured the elastic strain in circumferential direction $\epsilon_\theta$ for {\it Physcomitrium patens} caulonema tip cell (Fig.~\ref{fig:moss_image}), whose cell width is measured under normal conditions (i.e., turgid configuration) and with a treatment of 15$\%$ Mannitol in liquid media (i.e., relaxed configuration). The average circumferential elastic strain is around $15\%$ over 31 cell outlines across 6 different experiments. This experimental evidence suggests that the elastic strains in plant tip cells are in general large enough to be measured experimentally, thus the above formulae can be used to infer the elastic moduli when the elastic strains (or stretches) are available. In order to measure the elastic stretches at different locations along the cell wall outline and in both the circumferential and meridional directions, we need to apply microbeads along the cell wall and track their coordinates. In this work, we propose such an image-based approach and theoretically interrogate its property and feasibility.

%The bulk modulus $K_h$ describes the resistance to compression and extension, while the shear modulus $\mu_h$ measures the elastic shear stiffness as response to shear deformation, of the cell wall. So a larger $K$ or a larger $\mu$, respectively, indicates a greater force is required to compress or deform the cell, implying a lower compressibility or high rigidity. On the contrary, smaller values mean higher softness.\par

%We have shown how the material properties distribution along the cell outline can be related to the tensions and stretch ratios, using Eqs.\ref{continuous_K}\&\ref{continuous_mu}. The inference scheme for reconstructing the distribution of material properties along the cell outline, based on the assumption that a constant turgor pressure $P\neq 0$ is effected all over the cell, is presented below.

\section{The marker point based inference scheme}\label{sec:step_infer}
\subsection{Step-function inference scheme of elastic moduli}
Now, we formulate the step-function inference scheme based on Eqs.~(\ref{eq:continuous_K}) and (\ref{eq:continuous_mu}) that computes the elastic moduli as step-function distributions based on the discrete material marker-point coordinates along the cell wall outline in both the relaxed and turgid configurations. 
We discretize the turgid configuration $\vec{r}(s)$  into $N$ linear segments by $N+1$ marker points, i.e., $\vec{r}_i=(z_i,r_i)$ for $i=1,2\dotsc, N+1$. For each material point currently located at $\vec{r}_i=(z_i,r_i)$, its relaxed coordinate without turgor pressure is $\vec{r}^0_i=(z^0_i,r^0_i)$ (Fig.~\ref{wallMechanics}b). For each linear segment between the marker points $\vec{r}_i=(z_i,r_i)$ and $\vec{r}_{i+1}=(z_{i+1},r_{i+1})$ (which corresponds to the linear segment between $\vec{r}^0_i$ and $\vec{r}^0_{i+1}$ in the relaxed configuration), we can infer its elastic moduli as constants based on the formulae (\ref{eq:continuous_K}) and (\ref{eq:continuous_mu}) 
\begin{align} \label{eq:K_discrete}
K_h^{i+1/2}&=\frac{\sigma_s^{i+1/2}+\sigma_\theta^{i+1/2}}{2(\lambda_s^{i+1/2}\lambda_\theta^{i+1/2}-1)}\\ \label{eq:mu_discrete}
\mu_h^{i+1/2}&=\frac{\sigma_s^{i+1/2}-\sigma_\theta^{i+1/2}}{1/(\lambda_\theta^{i+1/2})^{2}-1/(\lambda_s^{i+1/2})^{2}}
\end{align}
where the elastic stretches for the linear segment are calculated by 
\begin{align}
&\lambda_s^{i+1/2}=l^{i+1/2}/l_0^{i+1/2},\\
&\lambda_\theta^{i+1/2}=r^{i+1/2}/r_{0}^{i+1/2},
\end{align}
where $l^{i+1/2}$ and $l^{i+1/2}_0$ are the lengths of the linear segment on the turgid and relaxed configurations, respectively, and $r^{i+1/2}=(r_i+r_{i+1})/2$ and $r_{0}^{i+1/2}=(r^0_i+r^0_{i+1})/2$ are the averaged local radii from the axis of symmetry. The step-function inference scheme for the tensions $\sigma_s^{i+1/2}$ and $\sigma_\theta^{i+1/2}$ for each linear segment are presented in our previous work \cite{chelladurai2020inferring} and the details are shown in Appendix~\ref{sec:S1}. From Eqs.~(\ref{eq:K_discrete}) and (\ref{eq:mu_discrete}), we can directly observe that the inferred bulk modulus might be less accurate where $\lambda_s\lambda_\theta\approx 1$ and the inferred shear modulus might be less accurate where $\lambda_s\approx\lambda_\theta$ (which will be discussed below).

\subsection{Perturbation analysis} 
Given the cell wall is sufficiently axially symmetric, the accuracy and robustness of the step-function inference scheme clearly rely on the reconstruction fidelity of the material marker point coordinates. While plant cells do not fluctuate as much as animal cells due to their cell-wall structure, the perturbation to the actual marker-point locations can still occur during imaging acquisition and processing, and the final registration of the marker point coordinates. To investigate the robustness of our inference scheme to the systemic noise, we perform perturbation analysis to quantify the errors of the inferred elastic moduli due to the perturbation to the actual locations. We consider an arbitrary perturbation $(\delta z^0_i,\delta r^0_i)$ (and $(\delta z_i,\delta r_i)$) in the positions of cell marker points in the relaxed (and turgid) configurations. Assuming these perturbations are small, we can obtain the leading-order approximation of the resulting relative errors in the local inferred step-function elastic moduli (see Appendix~\ref{sec:S2} for detailed derivation)
\begin{align} \label{eq:perturbK}
\Big(\frac{\delta K_h}{K_h}\Big)^{i+\frac{1}{2}}&=\Big(\frac{\sigma_s}{\sigma_s\!+\!\sigma_\theta}\frac{\delta\sigma_s}{\sigma_s}\!+\!\frac{\sigma_\theta}{\sigma_s\!+\!\sigma_\theta}\frac{\delta\sigma_\theta}{\sigma_\theta}\Big)^{i+\frac{1}{2}}
\!+\!\Big(\frac{\lambda_s\lambda_{\theta}}{1\!-\!\lambda_s\lambda_{\theta}}(\frac{\delta\lambda_s}{\lambda_s}\!+\!\frac{\delta\lambda_\theta}{\lambda_\theta})\Big)^{i+\frac{1}{2}}, \\
\label{eq:perturbMu}
\Big(\frac{\delta\mu_h}{\mu_h}\Big)^{i+\frac{1}{2}}&\!=\!\Big(\frac{\sigma_s}{\sigma_s\!-\!\sigma_\theta}\frac{\delta\sigma_s}{\sigma_s}\!+\!\frac{\sigma_\theta}{\sigma_\theta\!-\!\sigma_s}\frac{\delta\sigma_\theta}{\sigma_\theta}\Big)^{i+\frac{1}{2}}
\!+\!\Big(\frac{2\lambda_{\theta}^2}{\lambda_{\theta}^2\!-\!\lambda_s^2}\frac{\delta\lambda_s}{\lambda_s}\!+\!\frac{2\lambda_s^2}{\lambda_s^2\!-\!\lambda_{\theta}^2}\frac{\delta\lambda_\theta}{\lambda_\theta}\Big)^{i+\frac{1}{2}},
\end{align}
where the relative errors in tensions and elastic stretches are given in terms of $(\delta z^0_i,\delta r^0_i)$ and $(\delta z_i,\delta r_i)$ in Appendix~\ref{sec:S2}. 

The above results reveal the sensitivity of the inferred elastic moduli on the perturbations. We observe from Eq.~(\ref{eq:perturbK}) that the bulk modulus $K$ might suffer instability when $\lambda_s\lambda_\theta\approx 1$. It would be problematic if the turgor pressure is very small or the cell wall is too stiff such that the elastic strains are very small, i.e., $\epsilon_s=\lambda_s-1\approx 0$ and $\epsilon_\theta=\lambda_\theta-1\approx 0$. In fission yeasts \cite{abenza2015wall,atilgan2015morphogenesis,davi2019systematic} and our experiments in moss cells, the averaged elastic strain is about $10\%-20\%$, which supports the feasibility of our inference of $K$. On the other hand, the shear modulus $\mu$ in Eq.~(\ref{eq:perturbMu}) might suffer instability when $\sigma_s\approx\sigma_\theta$ and/or $\lambda_s\approx\lambda_\theta$, which may occur at the very tip of cell where the tensions and elastic stretches (or strains) are isotropic. It indicates that the sensitivity of the inferred shear modulus $\mu$ to noise can increase significantly near the tip. 

To further elucidate the effects of the perturbations on the inference scheme when the elastic strains ($\epsilon_s$ and $\epsilon_\theta$) are small, we rewrite Eqs.~(\ref{eq:perturbK}) and (\ref{eq:perturbMu}) in the form
%\begin{align} \label{eq:perturbK_strain}
%\Big(\frac{\delta K_h}{K_h}\Big)^{i+\frac{1}{2}}&=\Big(\frac{\sigma_s}{\sigma_s\!+\!\sigma_\theta}\frac{\delta\sigma_s}{\sigma_s}\!+\!\frac{\sigma_\theta}{\sigma_s\!+\!\sigma_\theta}\frac{\delta\sigma_\theta}{\sigma_\theta}\Big)^{i+\frac{1}{2}}\!+\!\Big(\frac{\epsilon_s}{\epsilon_s\!+\!\epsilon_\theta}\frac{\delta\epsilon_s}{\epsilon_s}\!+\!\frac{\epsilon_\theta}{\epsilon_s\!+\!\epsilon_\theta}\frac{\delta\epsilon_\theta}{\epsilon_\theta}\Big)^{i+\frac{1}{2}},\\ \label{eq:perturbMu_strain}
%\Big(\frac{\delta\mu_h}{\mu_h}\Big)^{i+\frac{1}{2}}&=\Big(\frac{\sigma_s}{\sigma_s\!-\!\sigma_\theta}\frac{\delta\sigma_s}{\sigma_s}\!+\!\frac{\sigma_\theta}{\sigma_\theta\!-\!\sigma_s}\frac{\delta\sigma_\theta}{\sigma_\theta}\Big)^{i+\frac{1}{2}}\!+\!\Big(\frac{\epsilon_s}{\epsilon_s\!-\!\epsilon_\theta}\frac{\delta\epsilon_s}{\epsilon_s}\!+\!\frac{\epsilon_\theta}{\epsilon_\theta\!-\!\epsilon_s}\frac{\delta\epsilon_\theta}{\epsilon_\theta}\Big)^{i+\frac{1}{2}}.
%\end{align}
\begin{align} \label{eq:perturbK_strain}
\Big(\frac{\delta K_h}{K_h}\Big)^{i+\frac{1}{2}}&=\Big(\frac{\sigma_s}{\sigma_s\!+\!\sigma_\theta}\frac{\delta\sigma_s}{\sigma_s}\!+\!\frac{\sigma_\theta}{\sigma_s\!+\!\sigma_\theta}\frac{\delta\sigma_\theta}{\sigma_\theta}\Big)^{i+\frac{1}{2}}\!-\!\Big(\frac{1+\epsilon_s+\epsilon_\theta}{\epsilon_s\!+\!\epsilon_\theta}(\frac{\delta\lambda_s}{\lambda_s}\!+\frac{\delta\lambda_\theta}{\lambda_\theta})\Big)^{i+\frac{1}{2}},\\ \label{eq:perturbMu_strain}
\Big(\frac{\delta\mu_h}{\mu_h}\Big)^{i+\frac{1}{2}}&=\Big(\frac{\sigma_s}{\sigma_s\!-\!\sigma_\theta}\frac{\delta\sigma_s}{\sigma_s}\!+\!\frac{\sigma_\theta}{\sigma_\theta\!-\!\sigma_s}\frac{\delta\sigma_\theta}{\sigma_\theta}\Big)^{i+\frac{1}{2}}\!+\!\Big(\frac{1+2\epsilon_\theta}{\epsilon_\theta\!-\!\epsilon_s}\frac{\delta\lambda_s}{\lambda_s}\!+\!\frac{1+2\epsilon_s}{\epsilon_s\!-\!\epsilon_\theta}\frac{\delta\lambda_\theta}{\lambda_\theta}\Big)^{i+\frac{1}{2}}.
\end{align}
where we assume that $\epsilon_s$ and $\epsilon_\theta$, though small, are still much larger than the perturbations in the marker-point coordinates $\delta z^0_i$, $\delta r^0_i$, $\delta z_i$ and $\delta r_i$. This is the case of our numerical simulations where $\epsilon_s$ and $\epsilon_\theta$ are around $10\%$ while the perturbations in the marker-point coordinates are about $1\%$ of the cell radius maximum (see Sec.~\ref{sec:primary_inf}). We can observe that the $\delta K/K$ and $\delta\mu/\mu$ depend on the relative errors in tensions and elastic stretches. We will see in next section that although the relative errors $(\delta \sigma_s/\sigma_s, \delta \sigma_\theta/\sigma_\theta, \delta \lambda_s/\lambda_s, \delta \lambda_\theta/\lambda_\theta)$ are moderately sensitive to noise, their effects on $\delta K/K$ and $\delta\mu/\mu$ can be amplified by their coefficients in Eqs.~(\ref{eq:perturbK_strain}) and (\ref{eq:perturbMu_strain}). When both $\epsilon_s$ and $\epsilon_\theta$ are positive, due to the subtraction in the denominators in Eq.~(\ref{eq:perturbMu_strain}), instead of summation in Eq.~(\ref{eq:perturbK_strain}), we can predict that the inferred bulk modulus $K$ will be more accurate than the inferred shear modulus $\mu$, which will be confirmed in Fig.~\ref{fig:perturbedInference} in the next section.  
 
At last, it is counter-intuitive but worth-noting that the inference scheme becomes more robust to noise when the discretization of cell outline is coarser, which has been shown both analytically and quantitatively in the previous work \cite{chelladurai2020inferring} (also see Appendix~\ref{sec:S2}). While equidistant marker points along the cell outline are not necessary in the proposed method and are even impractical in experiments, in the following numerical simulations we implement equidistant discretization for illustrative purposes. Since the focus of the analysis above is the effect of noise on the marker-point registration, the perturbations in turgor pressure $P$ are not considered here. 

\section{Primary inference of elastic moduli from synthetic data}\label{sec:primary_inf}
\subsection{Inferring elastic moduli from synthetic image data}
\begin{figure}[b]
	\centering
	\includegraphics[width=1\linewidth]{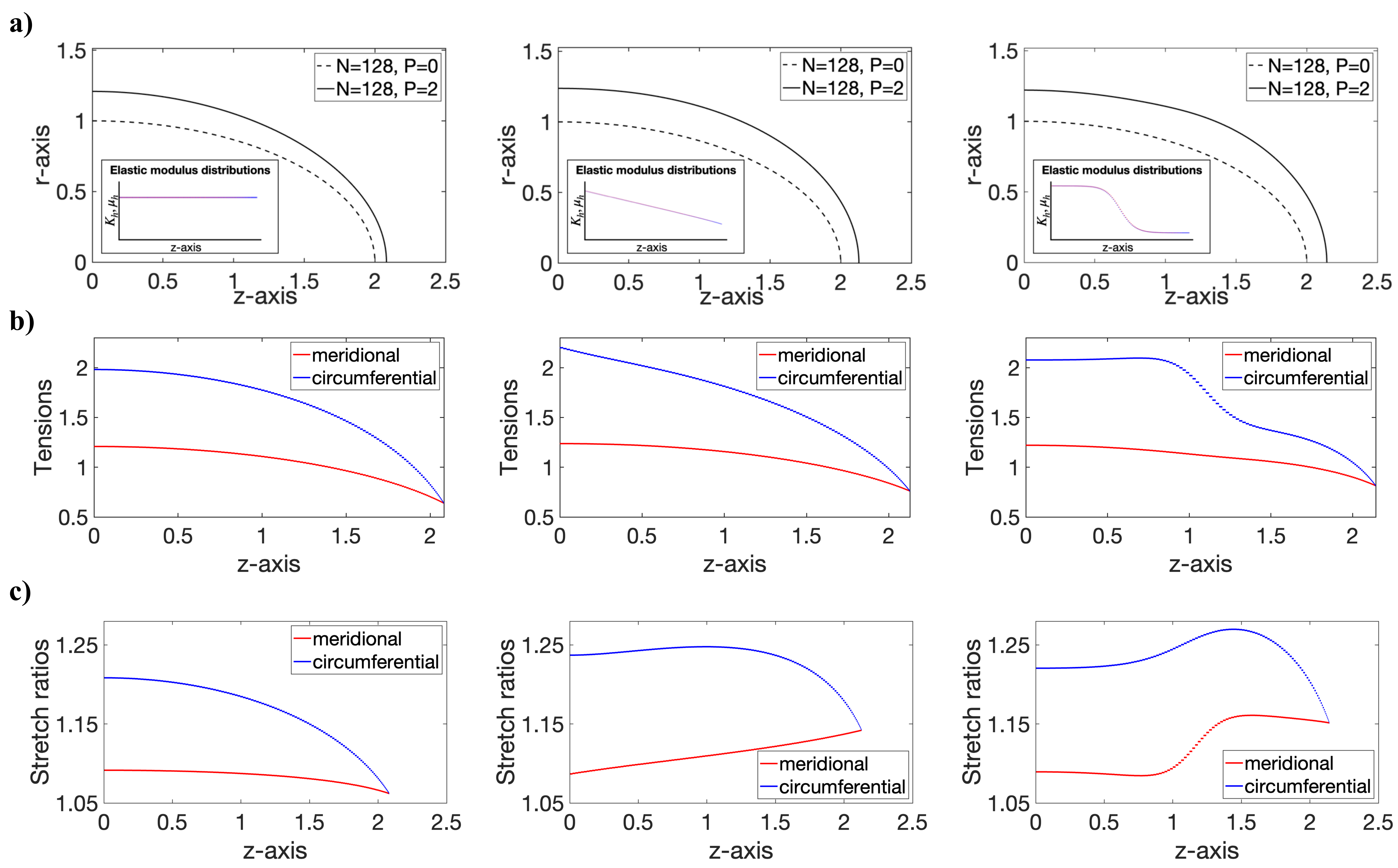}
	\caption{\textbf{Simulations of cell wall expansion and synthetic cell outline data.} \textbf{a)} Synthetic cell outlines without and with turgor pressure ($P=2$). By prescribing the shape of the cell outline without turgor pressure (dash lines), the turgid cell shape (solid lines) is computed from the simulation using $N=128$ linear segments, where three cases of cell-wall elastic modulus distributions (constant, linearly decreasing, nonlinearly decreasing) are considered (shown in the insets for each column). \textbf{b)} Meridional (red) and circumferential (blue) tension along the cell wall in the turgid configuration, computed from the simulations. \textbf{c)} Elastic stretch ratios $\lambda_s$ and $\lambda_\theta$ computed by comparing the turgid and relaxed cell outline profiles. Elastic strains can be computed by $\epsilon_s=\lambda_s-1$ and $\epsilon_\theta=\lambda_\theta-1$. 
	\label{fig:simulatedShapes}}
\end{figure}
First, we perform our step-function inference scheme for elastic moduli of the cell wall based on synthetic data of marker point coordinates along the cell wall with no perturbation. They are generated by a computational model of a hyper-elastic membrane deformation \cite{chelladurai2020inferring} with the following set up. The cell outline is assumed to be a half-ellipse \cite{fayant2010finite}, $r^2+z^2/4=1$, which mimics the shape of elongated plant tip cells. We discretize the cell outline into $N=128$ linear segments by $129$ marker points in our simulations. The turgor pressure is set to be $P=2$ to generate realistic elastic strains around $10\%-20\%$ found for fission yeast in \cite{atilgan2015morphogenesis} and our experiments for moss cells (see Fig.~\ref{fig:moss_image}). Previous works suggest that the cell wall is softer at the apical (tip) region than the shank (rear) region \cite{goriely2003self,fayant2010finite}. To test if our inference scheme is able to detect the existing gradient along the cell wall, we in particular consider three types of elastic modulus distributions along the cell wall: constant, linearly decreasing from the shank to tip, and nonlinearly decreasing from the shank to tip like a sigmoid curve \cite{fayant2010finite}. Without loss of generality, we set the bulk modulus and shear modulus as identical along the cell wall, i.e., $K_h=\mu_h$, which coincides with the case of Poisson ratio $\nu=0$ in linear elasticity used in \cite{atilgan2015morphogenesis} (see the discussion of the relation between nonlinear and linear elasticity constitutive laws in Sec.~\ref{sec:mechanics}). We implement the following three cases of synthetic elastic moduli for thin-shell elastic deformation : $K_h=\mu_h=5$, $K_h=\mu_h=5-1.25z^0$, and $K_h=\mu_h=1.25(1-\tanh[(z^0-1)/0.2])+2.5$ for $0\leq z^0\leq 2$, where the magnitude is chosen for generating appropriate elastic strains. However, notice that our inference does not treat $K_h=\mu_h$ as {\it a priori} knowledge and is applicable in cases when $K_h\neq\mu_h$.

We show the simulation results of the turgid configuration juxtaposed with the relaxed configuration in the three cases of elastic modulus distributions in Fig.~\ref{fig:simulatedShapes}a.  The corresponding tensions and elastic stretches computed from the simulation are shown in Fig.~\ref{fig:simulatedShapes}b and c. Comparing with the first case with constant elastic moduli, we observe that the wall surface expansion for the latter two cases is more pronounced near the tip region, due to the decreasing elastic moduli from the shank to the tip (Fig.~\ref{fig:simulatedShapes}a). In addition, the circumferential tensions and strains are greater than the meridional ones except at the tip where they are identical  (Fig.~\ref{fig:simulatedShapes}b). We also observe that the resulting circumferential elastic strains in cell width for all three cases are around $20\%$ at the rear boundary ($z=0$), close to the case for {\it Physcomitrium patens} caulonema tip cell (Fig.~\ref{fig:moss_image}). Interestingly, for the two latter cases with a decreasing gradient of elastic moduli from the shank to the tip,  both meridional and circumferential strains become larger near the tip region compared to the case with constant elastic moduli (Fig.~\ref{fig:simulatedShapes}c).

\begin{figure}[t]
	\centering
	\includegraphics[width=1\linewidth]{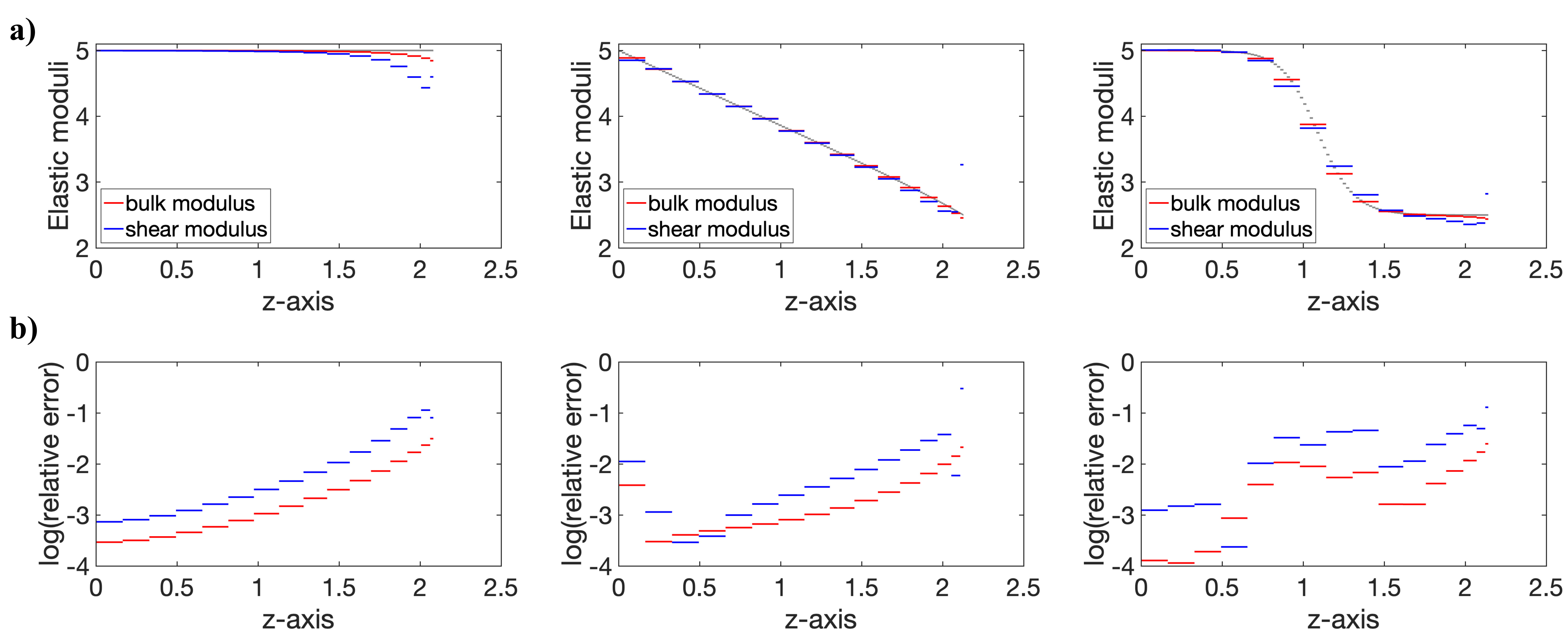}
	\caption{\textbf{Step-function inference of elastic moduli from synthetic data (without added noise). a)} The inference of the bulk modulus (red) and shear modulus (blue) with $N_I=16$ linear segments, juxtaposed with the prescribed underlying elastic moduli (grey) for the three different cases (three columns). \textbf{b)} The logarithm ($\log_{10}$) of the relative errors between the step-function inference and prescribed values for bulk modulus (red) and shear modulus (blue).
	\label{fig:noNoiseInference}}
\end{figure}
Using the synthetic cell outline coordinates without ($P=0$) and with ($P=2$) turgor pressure in Fig.~\ref{fig:simulatedShapes}a, we apply our step-function inference scheme (through Eqs.~(\ref{eq:K_discrete}) and (\ref{eq:mu_discrete})) with different numbers of equidistant marker points along the cell outline. In Fig.~\ref{fig:noNoiseInference}, we show the inferred elastic modulus distributions by using $N_I=16$ linear segments ($17$ marker points) and their relative errors with respect to the prescribed values. We find that the inference scheme gives accurate predictions of both moduli along the cell outline except near the tip for the shear modulus, as we have discussed in Sec.~\ref{sec:step_infer}.

\subsection{Inferring elastic moduli from noisy data}
\begin{figure}[t]
	\centering
	\includegraphics[width=1\linewidth]{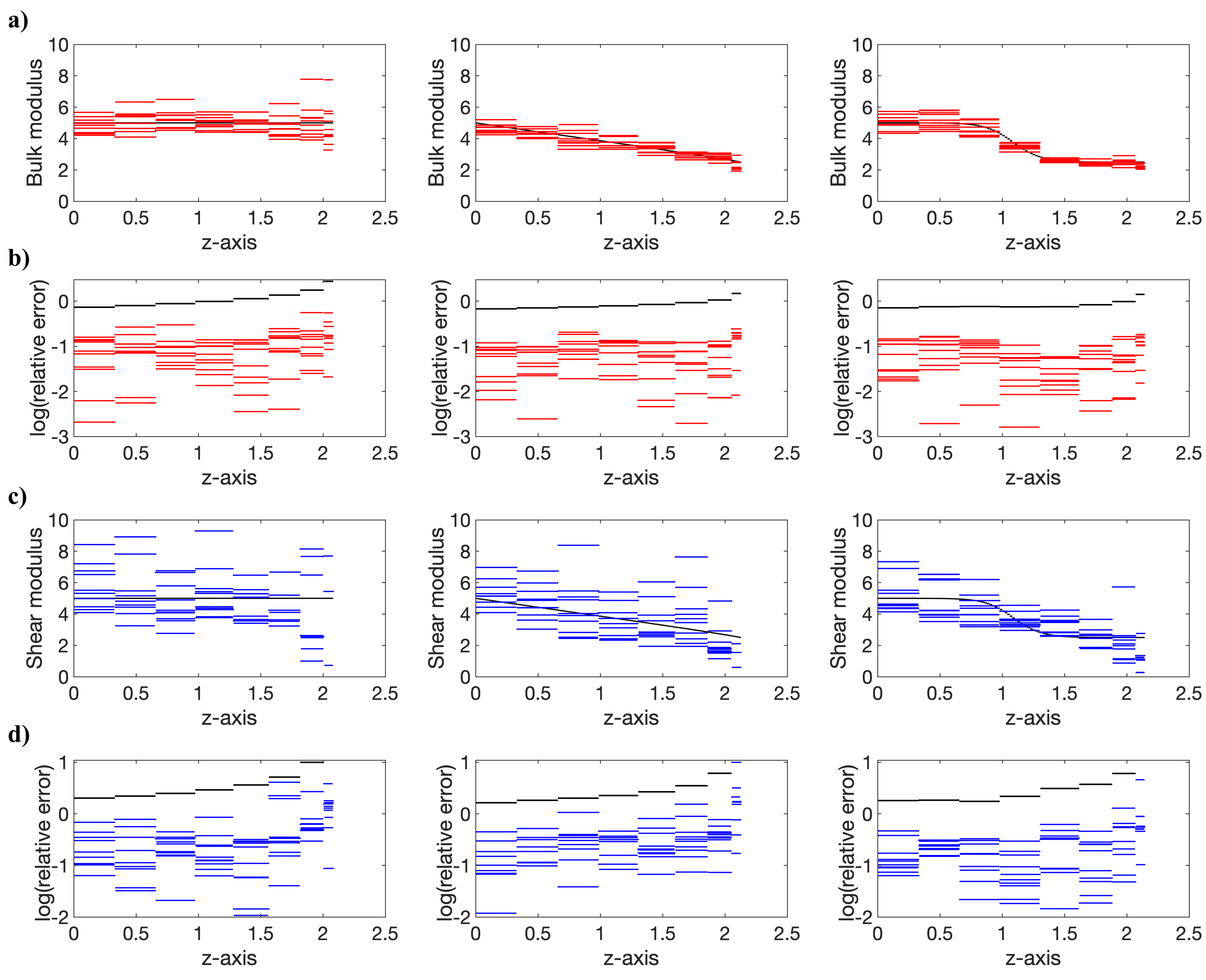}
	\caption{\textbf{Inference of elastic moduli from noisy data.} Step-function inference from noisy data for bulk modulus $\textbf{a)}$ and shear modulus $\textbf{c)}$ using $N_I=8$ linear segments. 10 step-function inferences based on 10 cell outlines with random noise are shown, where the noise are generated by adding random perturbations of $1\%\times$ maximal cell radius to the synthetic marker point positions along the cell wall. 
	The relative errors in inferred bulk modulus \textbf{b)} and shear modulus \textbf{d)} are plotted in base-10-logarithmic scale, compared with the estimate of error bounds (black) computed from the perturbation analysis. For better view, the rightmost error bound and some outliers of the inference are not shown for the shear modulus.
	\label{fig:perturbedInference}}
\end{figure}

To quantify the effects of noise in marker-point positions and validate the prediction by perturbation analysis, we perform the step-function inference scheme from noisy data. For each marker point $\vec{r}_i=(z_i,r_i)$ (and $\vec{r}^0_i=(z^0_i,r^0_i)$) from the synthetic cell outline, we displace its position by small random perturbations independently drawn from a uniform distribution $U(-\delta_m/2, \delta_m/2)$ such that $(z_i, r_i)\to (z_i+U(-\delta_m/2, \delta_m/2), r_i+U(-\delta_m/2, \delta_m/2))$, where $\delta_m$ defines as the noise magnitude setting the scale of the perturbation (and the same noise magnitude for $(z^0_i,r^0_i)$). Note that our analysis (see Appendix~\ref{sec:S2}) suggests that a coarser discretization with smaller $N_I$ is needed to keep the error in the inferred moduli effectively small when $\delta_m$ increases. We demonstrate that for our primary inference scheme with the displacement noise $\delta_m=1\%\times$ the maximal cell radius, $N_I=8$ linear segments along the cell outline (see Fig.~\ref{fig:perturbedInference}) is necessary to generate reliable results for the three cases of elastic modulus distributions, while $N_I=16$ is no longer providing reliable results (see below for details). The results from 10 cell outlines with random perturbations validate our perturbation analysis by showing that the errors from the numerical experiments are below the theoretical estimate of error bounds for both elastic moduli (Fig.~\ref{fig:perturbedInference}b and d). We observe that the primary inference for the bulk modulus is more accurate and the relative error is below $30\%$ except near the tip for all three cases of elastic modulus distributions, while the primary inference of the shear modulus is less accurate with relative errors between $10\%$ and $100\%$. This observation validates the conclusions from perturbation analysis that the inference is more sensitive to the perturbation near the tip and the inference of the shear modulus is more challenging than the bulk modulus. 

To further investigate how the perturbation on the marker-point locations is translated to the perturbation on the elastic moduli, we also show the intermediate inference results for tensions and stretch ratios in Fig.~\ref{fig:infTensions}, and find that the relative errors for all tensions and stretches are around $1\%$ and do not exceed $10\%$ even at the tip. This means that the error has been amplified from the intermediate to the final inference. We can observe from Eq.~(\ref{eq:perturbK_strain}) that the coefficient before the relative errors in stretches, i.e., $\frac{1+\epsilon_s+\epsilon_\theta}{\epsilon_s+\epsilon_\theta}$, is larger than 1 given that both $\epsilon_s$ and $\epsilon_\theta$ are small (see Fig.~\ref{fig:simulatedShapes}), and hence the relative errors in stretches are amplified in the inference of bulk modulus $K$. For the inference of shear modulus $\mu$ from Eq.~(\ref{eq:perturbMu_strain}), the relative errors in tensions can be amplified near the tip when $\sigma_s\approx \sigma_\theta$, while those in stretches are amplified more significantly along the whole cell outline with amplification coefficients $\frac{1+2\epsilon_\theta}{\epsilon_\theta\!-\!\epsilon_s}$ and $\frac{1+2\epsilon_s}{\epsilon_s\!-\!\epsilon_\theta}$ that can be as large as $\sim 7$ even at the rear boundary of the cell (see Fig.~\ref{fig:simulatedShapes}).
The booming inaccuracy in inferred elastic moduli demonstrates the difficulty of inference for elastic moduli compared to that of tensions and stretches due to this amplification effect. We also test the inference with finer resolution ($N_I=16$) and find that the relative errors could exceed $100\%$ for both moduli (see Fig.~\ref{fig:kmu_N16}), which demonstrates the higher sensitivity to perturbations for inference with high-resolution marker point spacing. To summarize, we find both analytically and numerically that the inference scheme is more robust to noise with coarser discretization, which, however, limits the resolution of the inferred elastic moduli because of its step-function representation.

\section{Inference of canonical elastic moduli using multiple cell data}\label{sec:canonical_infer}
In the previous work \cite{chelladurai2020inferring}, we proposed an optimization inference scheme for canonical tensions by fitting a polynomial with multiple cell data. We implemented this optimization inference scheme to measure the averaged tensions for two different cell types of moss protonemata (caulonemata and chloronemata), where the noise arises from both the image processing and the variation of cell-wall morphology from individual cells of the same type. Inspired by this optimization inference for tensions, we generalize this scheme for inferring the canonical elastic moduli distributions of cell walls for a specific cell type where single cell's morphologies in the relaxed and turgid configurations are due to perturbations from its canonical shape. We further improve this method by selecting the best-fitting polynomial among polynomials with degree of $1 \leq n_p \leq 15$, instead of asserting a degree of polynomial {\it a priori}. The upper limit of polynomial degree $15$ is chosen through a number of trials, where we have found that the accuracy of optimizing-polynomial inference scheme stops improving significantly with higher degree ($n_p>15$). The detailed mathematical model for finding an optimal polynomial inference by minimizing a cost function is presented in Appendix~\ref{sec:S3}. 
\begin{figure}[t]
	\centering
	\includegraphics[width=1\linewidth]{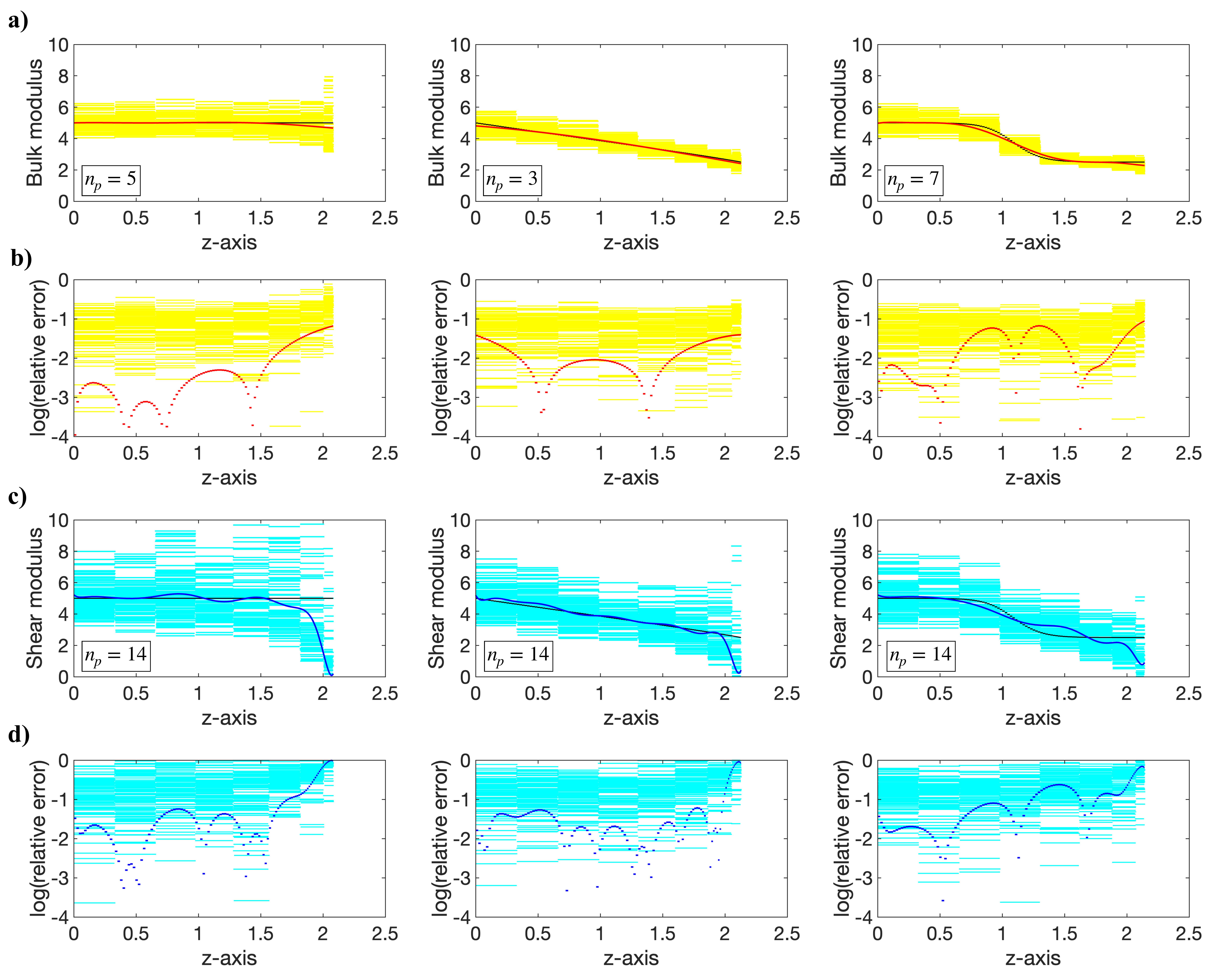}
	\caption{\textbf{Inference of canonical elastic moduli from multiple cell data.} 		
	The smooth-function inference of the bulk modulus $\textbf{a)}$ and shear modulus $\textbf{c)}$ are obtained by the optimization scheme (Approach 1) based on 100 synthetic cell outlines with $1\%$ random noise. For each cell outline, we use $N_I=8$ linear segments, and the optimal degree of polynomial fitting $n_p$ is shown in the insets. For bulk modulus, the smooth-function inference is shown in red, juxtaposed with the prescribed bulk modulus (black) and the $100$ primary step-function inferences (yellow). For shear modulus, the smooth-function inference is shown in blue, juxtaposed with the prescribed shear modulus (black) and the $100$ primary step-function inferences (cyan). The relative errors in inferred bulk modulus \textbf{b)} and shear modulus \textbf{d)} are plotted in $\log_{10}$ scale, where the color schemes are the same as $\textbf{a)}$ and $\textbf{c)}$. For better view, the rightmost error bound and some outliers of the inference are not shown for the shear modulus.
	\label{fig:optimizedInference}}
\end{figure}

Now we propose two different approaches: we can directly use the step-function inferences of elastic moduli based on multiple cell outlines to find a best-fitting polynomial inference for elastic moduli (called Approach 1 henceforth); or, we can first use the multiple cell-outline data to obtain step-function inferences for tensions and elastic stretches (using Eqs.~(\ref{eq:sigma_s_inf}-\ref{eq:lambda_theta_inf})), based on which we can find optimal polynomial representations for tensions and stretches, and then compute the elastic moduli by these polynomial tensions and stretches using Eqs.~(\ref{eq:continuous_K}) and (\ref{eq:continuous_mu}) (called Approach 2 henceforth).

The effectiveness of both approaches is tested using 100 samples of cell outline data with turgor pressure $P=2$ discretized with $N_I=8$ linear segments perturbed by noise with magnitude $\delta_m=1\%$. To avoid the distraction from extreme inaccuracies, we exclude some outliers, i.e., inferred values with more than three MAD (median absolute deviation) from the median, at each segment in the step function. We show the inference results by Approach 1 in Fig.~\ref{fig:optimizedInference} where the inset shows the degree of the best-fitting polynomial. For the bulk modulus, the optimal-polynomial inference almost overlaps with the prescribed values for all three cases of elastic modulus distributions, and the relative errors are significantly reduced, around $1\%$ or lower on average and not exceeding $10\%$ even near the tip. The inference for shear modulus is also significantly improved by the optimization scheme such that the relative errors on average are around $1\%-10\%$ for the first two cases and around $10\%-25\%$ for the sigmoid case, except near the tip. More remarkably, this optimization inference scheme is able to capture not only the general gradient (constant or decreasing) of the elastic moduli, but also some degree of nonlinearity (as shown by the inference for the sigmoid function case). We also test the scheme by Approach 2 (see Figs.~\ref{fig:optm_tensions} and \ref{fig:kmu_app2}) and there is no substantial differences between the two approaches. To summarize, both approaches based on the optimization scheme with multiple cell data remarkably improve the overall inference of elastic moduli and in particular provide very accurate predictions for bulk modulus and moderately good predictions for shear modulus with consistent gradient and some degree of nonlinearity (away from the very tip).

\section{A cost-effective inference scheme using single-cell data}\label{sec:shift-marker-scheme}
\begin{figure}[t]
	\centering
	\includegraphics[width=1\linewidth]{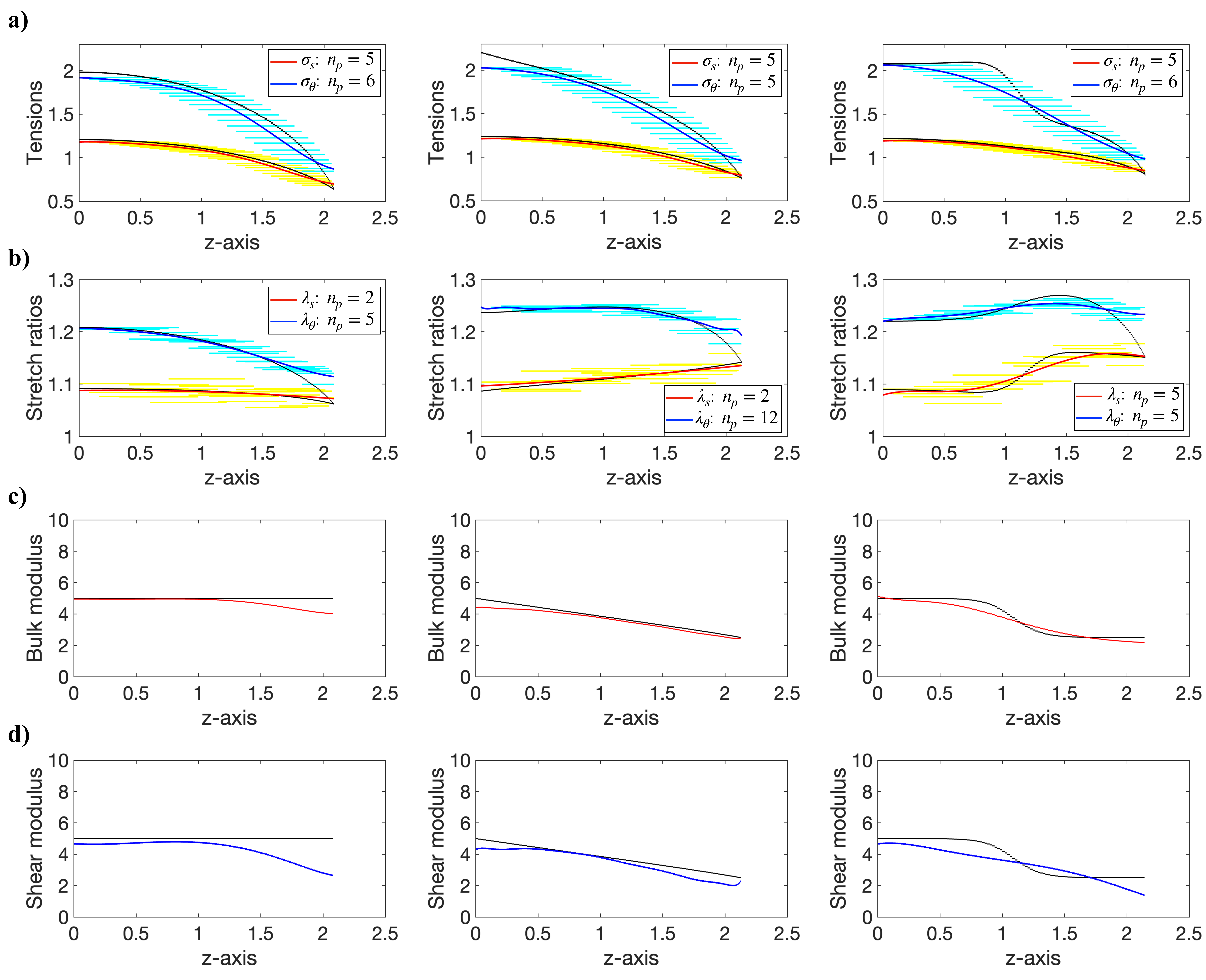}
	\caption{\textbf{The cost-effective inference on a single cell data with $1\%$ noise.}
	The smooth-function inferences of tensions \textbf{a)} and stretch ratios \textbf{b)} are obtained by fitting to the constant-value inferences (cyan and yellow), which are based on a set of 9 low-resolution data samples. See text for details. The optimal degrees of polynomial for the tension and elastic stretch inferences are shown in the insets. The smooth-function inferences for bulk modulus (red) \textbf{c)} and shear modulus (blue) \textbf{d)} are calculated using the polynomial inferences of tensions and stretches via Eqs.(\ref{eq:continuous_K}) and (\ref{eq:continuous_mu}), and are shown juxtaposed to the prescribed distributions (black). 
	\label{fig:shifting_markers}}
\end{figure}

Although the optimization inference scheme using multiple cell data provides high-precision predictions for elastic moduli, its feasibility and effectiveness might be limited by the availability of cell samples in practice. Therefore, we propose a cost-effective inference scheme from a single cell with high-resolution distribution of marker points. From the above results, we have shown that there exists a lower bound for the distance between two marker points in the relaxed configuration, $d$, to generate reliable inference results against noise with magnitude $\delta_m$. Given a single-cell outline in the relaxed configuration with total arclength $D$, we would only have $N_I$ reliable constant values from the step-function inference where $N_I\leq D/d$. For example, from the synthetic data  (Fig. \ref{fig:simulatedShapes}), $D$ is fixed by the half-ellipse $r^2+z^2/4=1$. If $d=32\Delta s^0$ (for stability consideration) where $\Delta s^0 = D/128$ with $129$ equidistant marker points, we would only have $N_I =4$ constant inferences among the $5$ marker points with indices $[1, 33, 65, 97, 129]$. The information of marker-point coordinates with other indices are not utilized. Since the success of the optimization schemes in the previous section relies on the abundance of constant-value inference data samples, we propose to increase the number of constant-value inferences within one cell by shifting the pair of marker points along the cell outline under a higher spatial resolution $\Delta N$. This resolution is determined by the maximum between the spacing between available marker points and the diffraction limit. For example, if $\Delta N= 4\Delta s^0$, we can generate $9$ sets of step-function samples where each set provides $3$ constant-value inferences. In detail, the $9$ sets of marker points (from the total $129$ marker points) can be grouped by their indices $[(1, 33, 65, 97), (5, 37, 69, 101),\ldots, (33, 65, 97, 129)]$, and a total number of $3\times 9=27$ constant-value inferences can be generated. Then, we can implement the optimization Approach 1 and Approach 2 to integrate the constant-value inferences into smooth function representations. Although two approaches give similar inference results, Approach 2 seems to be slightly better with less deflection at the tip. See Fig.~\ref{fig:shifting_markers} for the optimization results from Approach 2, where polynomial representation is implemented on the tensions and elastic stretches (Fig.~\ref{fig:shifting_markers} a,b), and the elastic moduli are computed through Eqs.(\ref{eq:continuous_K}) and (\ref{eq:continuous_mu}) (Fig.~\ref{fig:shifting_markers} c,d).  Interestingly, this cost-effective inference scheme using single cell data still provides high-precision predictions for elastic moduli away from the tip for the constant and linearly decreasing elastic moduli cases, and the result is almost as good as the inference by the optimization inference using multiple cell data (compare with Fig.~\ref{fig:optimizedInference}a,c). However, the cost-effective inference gives unrealistic gradient near the tip for the constant elastic moduli case. For the sigmoid-function case, the cost-effective inference cannot capture the nonlinear transition of the elastic moduli but can still infer the general decreasing trend. To check if the change of shifting-step size $\Delta N$ would improve the results, we have further tried $\Delta N= 8\Delta s^0$ (or $\Delta N= \Delta s^0$), which groups the marker points by $[(1, 33, 65, 97), (9, 41, 73, 105), \ldots, (33, 65, 97, 129)]$ (or $[(1, 33, 65, 97), (2, 34, 66, 98), \ldots, (33, 65, 97, 129)]$) with a total number of $3\times5=15$ (or $3\times33=99$) constant inferences. The results looks similar (see Fig.~\ref{fig:low_shifting} for the case $\Delta N= 8\Delta s^0$ for example). However, we find further increasing $\Delta N$ beyond $\Delta N= 8\Delta s^0$ starts to worsen the inference results due to the lack of constant-value inferences. Overall, the cost-effective inference scheme only requires a single cell data with high resolution such that enough number of constant-value inferences can be obtained. It can predict the consistent linear trend of elastic moduli (away from the tip) but not the nonlinearity. 

\section{Discussion and Conclusion}\label{sec:discussion}
In this paper, we have devised an image-based inference framework to compute the distribution of elastic moduli along the cell-wall surface. We have considered cells that maintain an axial symmetric geometry and the wall elasticity is transversely isotropic with respect to the wall thickness dimension. Under this consideration, introducing two elastic moduli is sufficient to describe the elastic property of the cell wall in the regime of small elastic strains (i.e., linear elasticity). We have further assumed the mechanical equilibrium of the thin cell wall is dominated by the interaction between turgor pressure and the lateral tensions due to wall-surface stretching, which allows us to measure the distribution of tensions using the marker-point coordinates along the cell outline through Young-Laplace Law. By tracking marker-point locations along the cell outline in the turgid and relaxed configurations, we can further measure the distributions of elastic stretches (or equivalently the elastic strains). Given tensions and strains, we can finally compute the distribution of the two elastic moduli algebraically.

Inspired by the nonlinear stress-strain relation of cell wall under lateral stretching \cite{zhang2021molecular}, we have proposed a hyper-elastic constitutive law \cite{wei2021eulerian} instead of Hooke's Law \cite{abenza2015wall,atilgan2015morphogenesis,davi2019systematic}. The constitutive law is chosen for a proof of concept, and whether it is an effective approximation for the cell wall mechanics needs further experimental validation in tip-growing cells such as moss protonemata. As long as the cell wall behaves like a transversely isotropic material under lateral tension, this framework can be easily extended to accommodate other constitutive laws with two elastic constants. In tip growing cells where it has been analyzed, such as root hairs \cite{o1954electron,dawes1959light,newcomb1965cytoplasmic} and moss protonemata \cite{wyatt2008cell,roberts2012moss}, the cellulose microfibril orientation appears to be random at the growing tip and its vicinity; this supports an isotropic composition, which is consistent with our framework. Within the current framework, we have analyzed how the marker-point spacing and location affect the stability of the inference scheme against the noise from image taking and processing. Similar to our previous work \cite{chelladurai2020inferring}, we have found that the inference scheme is less stable approaching the cell wall apex, and a coarser spacing between marker points can improve stability systemically.

Previous elastic moduli inferences \cite{abenza2015wall,atilgan2015morphogenesis,davi2019systematic} only interrogated the relative elastic moduli among discrete locations. To recover the true magnitudes of the moduli of the corresponding location, the values of cell wall thickness and turgor pressure need to be obtained or estimated. Given that previous theoretical modeling suggests that there is a continuous gradient of elastic moduli along the cell wall outline \cite{fayant2010finite,goriely2003self}, we have tested with noise if our inference algorithms are able to capture the existing gradient by inferring the underlying surface elastic moduli from the cell outline data synthesized by a computational model of hyper-elastic membrane deformation. From our first algorithm which integrates data from multiple ``virtual cells"  into two canonical distributions, we have found the bulk modulus distributions have been faithfully recovered globally in the cases of no gradient, a linear gradient, and a nonlinear gradient. Due to the lack of strain anisotropy close to the cell wall apex, the inferred shear modulus distribution is less accurate at the tip end, and can only capture the linear gradient (i.e., first-order approximation). We further developed a cost-effective algorithm which generates distribution of elastic moduli from a single `` virtual cell". This algorithm is able to capture the underlying elastic modulus distributions up to the first order. In the future, we will apply both algorithms on the cell wall of moss {\it Physcomitrium patens} and connect the distributions between elastic moduli and the cell wall composition among two cell types in the averaged sense (i.e., caulonema and chloronema) and within individual cells.

\backmatter

%\bmhead{Supplementary information}

\bmhead{Acknowledgments}
M.W. acknowledge partial funding from the National Science Foundation- Division of Mathematical Sciences (NSF-DMS) grant DMS-2144372 and DMS-2012330. L.V. acknowledges partial funding from the Division of Molecular and Cellular Biosciences (NSF-MCB) grant MCB-1253444. 
\bmhead{Data Availability Statements}
All data generated or analysed during this study are included in this published article.

%\section*{Declarations}

\begin{appendices}
\section{Experimental protocols and cell width measurements}\label{sec:S0}
Cell culture and imaging: protonemal filaments of {\it Physcomitrium patens} were cultured on WPi solid medium (Macronutrients: 1 mM  ${\text{MgSO}_4}$, 1mM $\text{Ca(NO}_3\text{)}_2$, 4 mM $\text{KNO}_3$, 89 µM Fe-EDTA,  1.84 mM $\text{KH}_2\text{PO}_4$. See \cite{liu2011efficient} for micronutrient details.) plates at 25°C under a cycle of light of 8h dark and 16h day. After being ground in a homogenizer (OMNI International), the plants were sieved through a 70 µm Nylon cell strainer (BD Falcon) to obtain individual cells. After 7 days, the plants were added to liquid WPi before being transferred to microfluidic chambers (see section below). The chambers were then flowed with liquid WPi and stained with Direct Yellow 96 dye (AK Scientific) at a final concentration of 1 µg/mL to mark the cell wall \cite{ursache2018protocol}. An average of 5-10 caulonema tip cells were observed using a motorized stage on an Inverted Microscope (Axio Observer Zeiss) with a 40x lens NA 0.95 (40ms and 60ms exposure for brightfield and EGFP, respectively). A Z-series of 20 images were taken every 1 µm with each filter before and immediately after a treatment of 1 mL of a 15\% (w/v) mannitol solution in liquid WPi. 
 
Microfluidic chambers: microfluidic chambers were made using a PDMS solution (10:1 crosslinker) casted over 50 µm - 100 µm tall scotch tape molds (311+ 3M). After being cut out of the mold, two 1 mm holes were punched onto either side of the chamber. The chambers were then plasma bonded to 35 mm petri dishes with a glass bottom (MatTek). Ten mL syringes were attached to each side of the chamber through a 3-way Luer valve and two 25 cm flexible plastic tubes (0.02 inches inner diameter), where the valve allowed for additional syringes to be attached to the side so flow could be redirected.  
 
Cell-width measurements: measurements were made in ImageJ. A maximum Z projection of the EGFP channel was first acquired, and then the brightness profile was averaged over 12 µm for a length of 20 µm - 25 µm across the cell (see Fig.\ref{fig:moss_image}), at a region ~10-20 µm from the tip. The maximal peaks marking the wall outline were then used to calculate the width of the cell before and after treatment.

\section{Step-function inference scheme for tensions}\label{sec:S1}
\begin{figure}[t]
	\centering
	\includegraphics[width=1\linewidth]{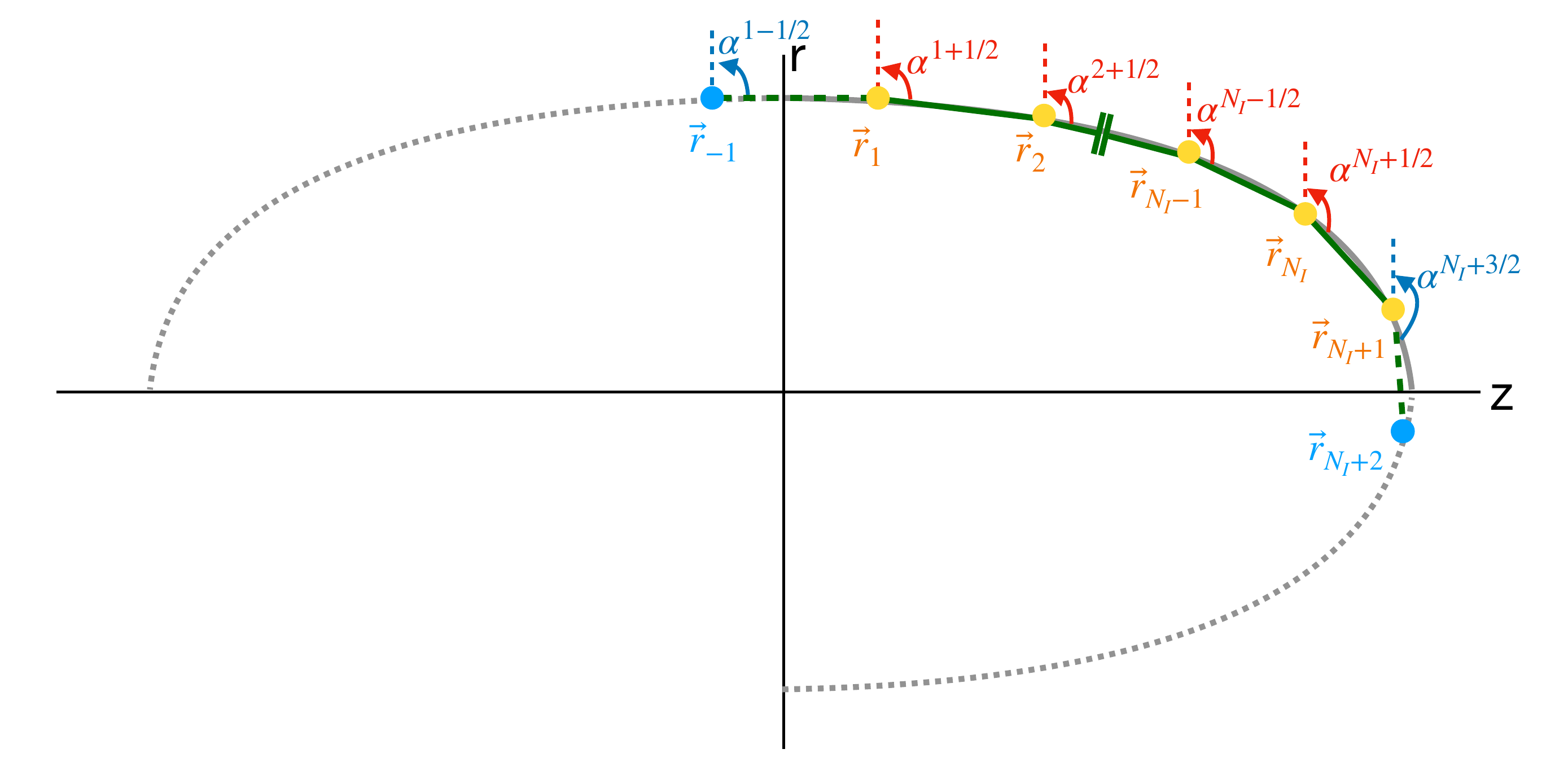}
	\caption{\textbf{Schematic of cell outline with ghost markers.} The cell wall (grey) is represented by the curve where the solid part represents the actual cell outline and dashed part represents the imaginary symmetrical shape. The actual marker points $\vec{r}_i$ ($i=1,2,\ldots,N_I+1$) are distributed along the actual cell outline while the ghost markers $\vec{r}_{-1}$ and $\vec{r}_{N_I+2}$ are located near the boundaries keeping the same arclength) distance between markers. The corresponding angles for all linear segments are indicated. 
	\label{fig:ghostPoints}}
\end{figure}

We discretize the turgid configuration $\vec{r}(s)=(z(s), r(s))$ (and relaxed configuration $\vec{r}^0(s^0)=(z^0(s^0), r^0(s^0))$) into $N$ linear segments by $N+1$ marker points, such that the marker points $\vec{r}_i=(z_i,r_i)$ and $\vec{r}^0_i=(z^0_i,r^0_i)$ (for $i=1,2, \dotsc, N+1$) correspond to the same material point in the initial relaxed and final turgid configurations, respectively. For each linear segment between $\vec{r}_i=(z_i,r_i)$ and $\vec{r}_{i+1}=(z_{i+1},r_{i+1})$, we define its length, average radius from the axis of symmetry and the angle with the direction of $r$-axis by
\begin{align}\label{eq:curv_s_inf}
&l^{i+1/2}=\sqrt{(z_{i+1}-z_i)^2+(r_{i+1}-r_i)^2},\\
&r^{i+1/2}=(r_i+r_{i+1})/2,\\
&\alpha^{i+1/2}=\arctan[(z_{i+1}-z_i)/(r_{i+1}-r_i)],
\end{align} 
The circumferential and meridional curvatures for each linear segment can be reconstructed by
\begin{align}\label{eq:curv_theta_inf}
&\kappa_\theta^{i+1/2}=\sin\alpha^{i+1/2}/r_m^{i+1/2},\\
\label{eq:curv_s_inf}
&\kappa_s^{i+1/2}=(\alpha^{i+3/2}-\alpha^{i-1/2})/(2l^{i+1/2}),
\end{align}
where $(\alpha^{i+3/2}-\alpha^{i-1/2})/2$ is the averaged angle rotated through the linear segment. In particular, for the two boundary linear segments, we define the angles $\alpha^{1-1/2}$ and $\alpha^{N+3/2}$ by using a ghost point technique where the ghost points are placed near the boundaries such that the arclength between markers are kept the same (see Fig.~\ref{fig:ghostPoints}). In particular, when the ghost point and its adjacent marker point are symmetric about $r$-axis or $z$-axis, we define $\alpha^{1-1/2}=\pi-\alpha^{1+1/2}$ and $\alpha^{N+3/2}=2\pi-\alpha^{N+1/2}$, respectively.

With the curvatures and measured pressure $P$, the tensions for each linear segments are computed as
\begin{align}\label{eq:sigma_s_inf}
\sigma_s^{i+1/2}&=P/(2\kappa_\theta^{i+1/2}),\\ \label{eq:sigma_theta_inf}
\sigma_\theta^{i+1/2}&=\sigma_s^{i+1/2}\times(2-\kappa_s^{i+1/2}/\kappa_\theta^{i+1/2}).
\end{align}
Defining the length $l_0^{i+1/2}$ and local radius $r_{0}^{i+1/2}$ for the initial relaxed configuration similarly for the turgid configuration, the elastic stretch ratios are calculated by
\begin{align}\label{eq:lambda_s_inf}
\lambda_s^{i+1/2}&=l^{i+1/2}/l_0^{i+1/2}\\ \label{eq:lambda_theta_inf}
\lambda_\theta^{i+1/2}&=r^{i+1/2}/r_{0}^{i+1/2}
\end{align}
Finally, based on the tensions and stretch ratios, the elastic moduli can be inferred by the formula
\begin{align}
K_h^{i+1/2}&=\frac{\sigma_s^{i+1/2}+\sigma_\theta^{i+1/2}}{2(\lambda_s^{i+1/2}\lambda_\theta^{i+1/2}-1)}\\
\mu_h^{i+1/2}&=\frac{\sigma_s^{i+1/2}-\sigma_\theta^{i+1/2}}{1/(\lambda_\theta^{i+1/2})^{2}-1/(\lambda_s^{i+1/2})^{2}}\label{mu_inference}
\end{align}

\section{Perturbation analysis of tensions and stretch ratios}\label{sec:S2}
Assume the position of each marker point on the turgid cell outline $(z_i, r_i)$ is displaced by small perturbations $(\delta z_i, \delta r_i)$. The previous work \cite{chelladurai2020inferring} has shown the leading order of relative errors in tensions due to the position perturbations
\begin{align}
(\frac{\delta\sigma_s}{\sigma_s})^{i+1/2}&=-(\frac{\delta\kappa_\theta}{\kappa_\theta})^{i+1/2}\label{perturbTensM},\\
(\frac{\delta\sigma_\theta}{\sigma_\theta})^{i+1/2}&=-(\frac{2(1-\beta)}{2-\beta}\frac{\delta\kappa_\theta}{\kappa_\theta}+\frac{\beta}{2-\beta}\frac{\delta\kappa_s}{\kappa_s})^{i+1/2},\label{perturbTensC}
\end{align}
where $\beta=\kappa_s/\kappa_\theta$. The relative errors in the curvatures are 
\begin{align}
(\frac{\delta\kappa_s}{\kappa_s})^{i+1/2}&=\frac{\delta\alpha^{i+3/2}-\delta\alpha^{i-1/2}}{\alpha^{i+3/2}-\alpha^{i-1/2}}-(\frac{\delta l}{l})^{i+1/2},\label{perturbCurvM}\\
(\frac{\delta\kappa_\theta}{\kappa_\theta})^{i+1/2}&=\frac{\delta\alpha^{i+1/2}}{\tan\alpha^{i+1/2}}-(\frac{\delta r}{r})^{i+1/2}\label{perturbCurvC},
\end{align}
where the relative error in the angle is given by
\begin{equation}\label{perturbAngle}
\delta\alpha^{i+1/2}=\frac{\tan\alpha^{i+1/2}}{1+(\tan\alpha^{i+1/2})^2}(\frac{\delta z_{i+1}-\delta z_i}{z_{i+1}-z_i}-\frac{\delta r_{i+1}-\delta r_i}{r_{i+1}-r_i}).
\end{equation}
Note that the relative error $(\delta\kappa_s/\kappa_s)^{i+1/2}$ in Eq.~(\ref{perturbCurvM}) involves $\delta\alpha^{i+3/2}$ and $\delta\alpha^{i-1/2}$ and here we use the ghost point technic (see Fig.~\ref{fig:ghostPoints}) for computing $\delta\alpha^{1-1/2}$ and $\delta\alpha^{N+3/2}$. Moreover, when two marker points have $z_i=z_{i+1}$ or $r_i=r_{i+1}$, for instance the ghost marker point and its adjacent marker point are symmetrical to the $z$-axis or $r$-axis, the denominators in Eq.~(\ref{perturbAngle}) could be zero. Therefore, we provide a modified formula for the angle error in these cases by taking the limit $z_{i+1}-z_i\to 0$ or $r_{i+1}-r_i\to 0$
\begin{align*}
\delta\alpha^{i+1/2}&=\frac{\delta z_{i+1}-\delta z_i}{r_{i+1}-r_i} \quad \text{when}\quad z_{i+1}=z_i, \\
\delta\alpha^{i+1/2}&=\frac{\delta r_{i+1}-\delta r_i}{z_{i+1}-z_i} \quad \text{when}\quad r_{i+1}=r_i.
\end{align*}

The effects of perturbations on the inference of tensions have been discussed in \cite{chelladurai2020inferring} and can be summarized as follows. One can observe directly from Eq.~(\ref{perturbTensC}) that the inference of the circumferential tension might suffer instability at locations where $\beta=\kappa_s/\kappa_\theta\approx 2$. From Eq.~(\ref{perturbCurvC}), we can see that both tensions are very sensitive to perturbations near the tip where $\tan(\alpha^{i+1/2})\approx 0$ and $r^{i+1/2}\approx 0$. More importantly, Eq.~(\ref{perturbAngle}) indicates a counter-intuitive but  worth-noting conclusion that the inference scheme is less robust when the discretization is finer since the denominator $(z_{i+1}-z_i)\approx 0$ and $(r_{i+1}-r_i)\approx 0$ with high-resolution cell outline data.

Assuming the initial relaxed cell outline $(z^0_i, r^0_i)$ perturbed by noise $(\delta z^0_i, \delta r^0_i)$, we can compute the relative errors in stretch ratios 
\begin{align}
(\frac{\delta\lambda_s}{\lambda_s})^{i+1/2}&=(\frac{\delta l}{l})^{i+1/2}-(\frac{\delta l_0}{l_0})^{i+1/2}, \label{perturbStrM}\\
(\frac{\delta\lambda_\theta}{\lambda_\theta})^{i+1/2}&=(\frac{\delta r}{r})^{i+1/2}-(\frac{\delta r_{0}}{r_{0}})^{i+1/2},\label{perturbStrC}
\end{align}
where the errors in length $l^{i+1/2}$ and average radius $r^{i+1/2}$ are 
\begin{align}
\delta l^{i+1/2}&=\sin\alpha^{i+1/2}(\delta z_{i+1}-\delta z_i)+\cos\alpha^{i+1/2}(\delta r_{i+1}-\delta r_i),\label{perturbL}\\
\delta r^{i+1/2}&=(\delta r_{i+1}+\delta r_i)/2 \label{perturbR}.
\end{align}
The errors $\delta l_0^{i+1/2}$ and $\delta r_0^{i+1/2}$ in the relaxed configuration can be obtained similarly. Then we can obtain the relative errors in the elastic moduli in terms of relative errors in tensions and stretch ratios
\begin{align*}
\Big(\frac{\delta K_h}{K_h}\Big)^{i+\frac{1}{2}}&=\Big(\frac{\sigma_s}{\sigma_s\!+\!\sigma_\theta}\frac{\delta\sigma_s}{\sigma_s}\!+\!\frac{\sigma_\theta}{\sigma_s\!+\!\sigma_\theta}\frac{\delta\sigma_\theta}{\sigma_\theta}\Big)^{i+\frac{1}{2}}
\!+\!\Big(\frac{\lambda_s\lambda_{\theta}}{1\!-\!\lambda_s\lambda_{\theta}}(\frac{\delta\lambda_s}{\lambda_s}\!+\!\frac{\delta\lambda_\theta}{\lambda_\theta})\Big)^{i+\frac{1}{2}}, \\
\Big(\frac{\delta\mu_h}{\mu_h}\Big)^{i+\frac{1}{2}}&\!=\!\Big(\frac{\sigma_s}{\sigma_s\!-\!\sigma_\theta}\frac{\delta\sigma_s}{\sigma_s}\!+\!\frac{\sigma_\theta}{\sigma_\theta\!-\!\sigma_s}\frac{\delta\sigma_\theta}{\sigma_\theta}\Big)^{i+\frac{1}{2}}
\!+\!\Big(\frac{2\lambda_{\theta}^2}{\lambda_{\theta}^2\!-\!\lambda_s^2}\frac{\delta\lambda_s}{\lambda_s}\!+\!\frac{2\lambda_s^2}{\lambda_s^2\!-\!\lambda_{\theta}^2}\frac{\delta\lambda_\theta}{\lambda_\theta}\Big)^{i+\frac{1}{2}}.
\end{align*}

%Likewise, for the elastic stretch ratios, Eqs.\ref{perturbStrM} \& \ref{perturbStrC} show their overall high sensitivity to perturbations as markers become closer to each other along the $z$- and $r$-axis. Although counterintuitive, the observations make the conclusion that the inference scheme has higher (lower) robustness to noise with coarser (finer) discretization.\par

\section{Inference scheme using multiple cell data}\label{sec:S3}
To find an optimal-polynomial inference based on multiple cell data, we consider the inference for an interested quantity (such as tensions, elastic stretches and elastic moduli) in the form of a polynomial of coordinate $z$ of degree $n_p$
\begin{align*}
P(z) \, = \sum_{i=0}^{n_p}a_i z^i,
\end{align*}
in which $a_i$'s are the coefficients to be determined through minimization of the sum of the squared cost integral $\sum_{h=1}^{M} \int_{s_h} (f_h(s)\, -\, P(s))^2\, ds$ along the (discrete) cell outline $s_h$, over $M$ data samples, where $f_h(s)$ is the step-function inference based on the $h$-th cell outline data. Converting the cost integral along coordinate $z$, we obtain 
\begin{align}\label{cost_func}
\phi = \sum_{h=1}^{M}\int_{a_h}^{b_h}(f_h(z) - P(z))^2 W_h(z)dz
\end{align}
where $a_h$ and $b_h$ are the lower and upper bounds of $z$ in the discrete cell outline $s_h$, and $W_h(z)\, =\, ds/dz\, =\, 1/ \sin (\alpha (z))$ is the weight function. Differentiating Eq.~(\ref{cost_func}) with respect to the undetermined coefficients gives 
\begin{align*}
0=\frac{\partial \phi}{\partial a_k}&=-2\sum_{h=1}^{M}\int_{a_h}^{b_h}z^k f_h(z)W_h(z)dz+\sum_{h=1}^{M}\sum_{i=0}^{n_p}a_i\int_{a_h}^{b_h}z^{i+k}W_h(z)dz,\\
&+\sum_{h=1}^{M}\sum_{j=0}^{n_p}a_j \int_{a_h}^{b_h}z^{j+k}W_h(z)dz
\end{align*}
for $k=0,1, ..., n_p$. Rearranging the equation provides the optimal condition of minimization
\begin{align*}
\sum_{j=0}^{n_p}a_j \sum_{h=1}^{M} \int_{a_h}^{b_h}z^{i+j}W_h(z)dz =\sum_{h=1}^{M}\int_{a_h}^{b_h}z^i f_h(z)W_h(z)dz,
\end{align*}
which can be written in a linear system $Ha=c$ for solving the coefficients $a=(a_0, ..., a_{n_p})^T$ where
\begin{align*}
H_{i,j}=\sum_{h=1}^{M}\int_{a_h}^{b_h}z^{i+j}W_h(z)dz \qquad \text{and} \qquad c_{i}=\sum_{h=1}^{M}\int_{a_h}^{b_h}z^i f_h(z)W_h(z)dz
\end{align*}
for $i=0$, 1, ..., $n_p$ and $j=$0, 1, ..., $n_p$.

\section{Additional inference results from noisy data}\label{sec:S4}
%\subsubsection{Tensions and stretch ratios}

\begin{figure}[h]
	\centering
	\includegraphics[width=1\linewidth]{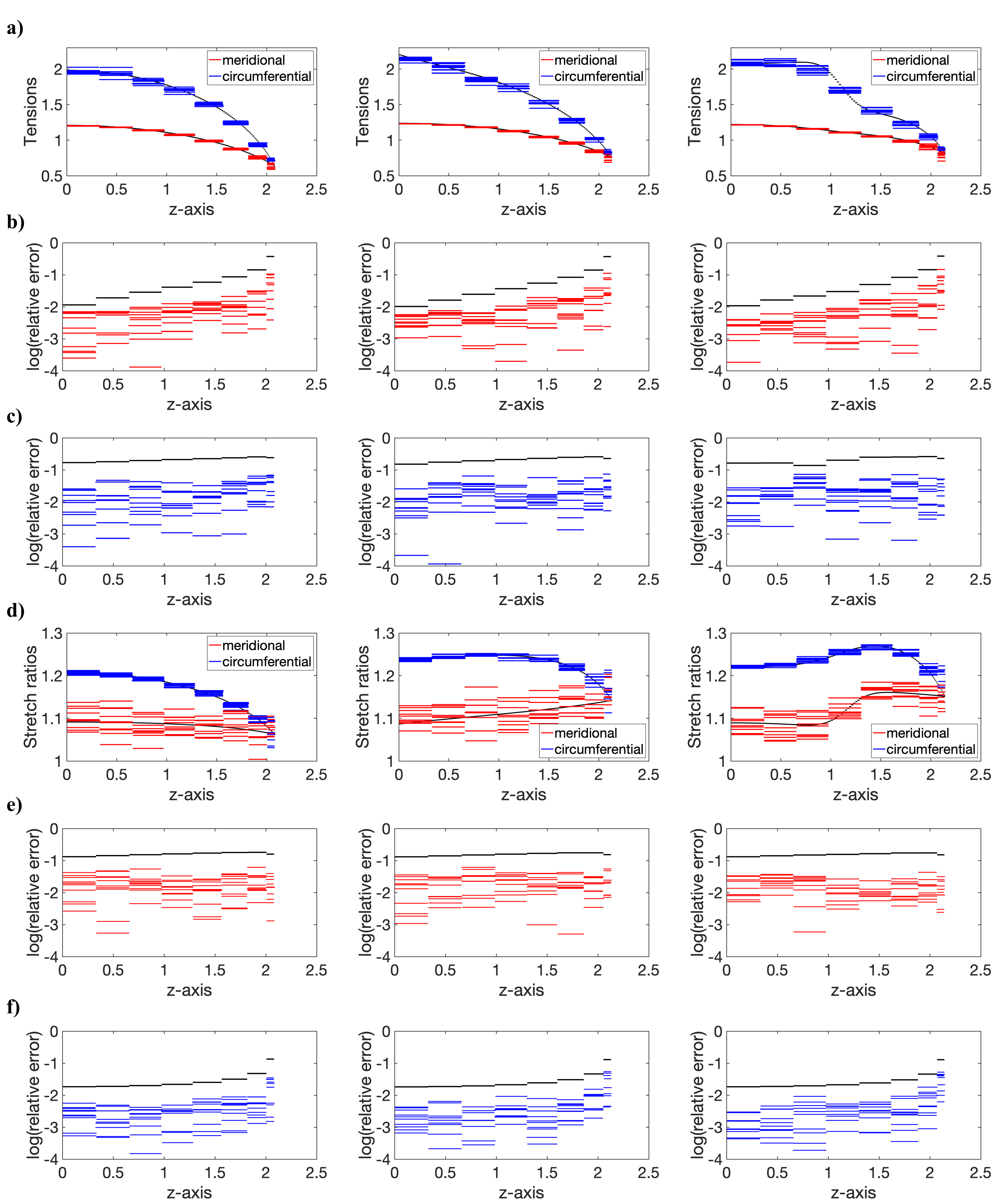}
	\caption{\textbf{Inference of tensions and stretch ratios from noisy data.} Step-function inferences for tensions $\textbf{a)}$ and stretch ratios $\textbf{d)}$ are inferred from noisy cell outline data with $1\%$ perturbations using $N_I=8$ linear segments. Ten inferences based on different noisy data are shown for three cases of cell-wall elastic modulus distributions (constant, linearly decreasing, nonlinearly decreasing).
	The relative errors in inferred meridional and circumferential tensions \textbf{b)} and \textbf{c)} and stretches \textbf{e)} and \textbf{f)} are plotted in $\log_{10}$ scale, juxtaposed with the estimate of error bounds (black) based on the perturbation analysis.
	\label{fig:infTensions}}
\end{figure}

\begin{figure}[h]
	\centering
	\includegraphics[width=1\linewidth]{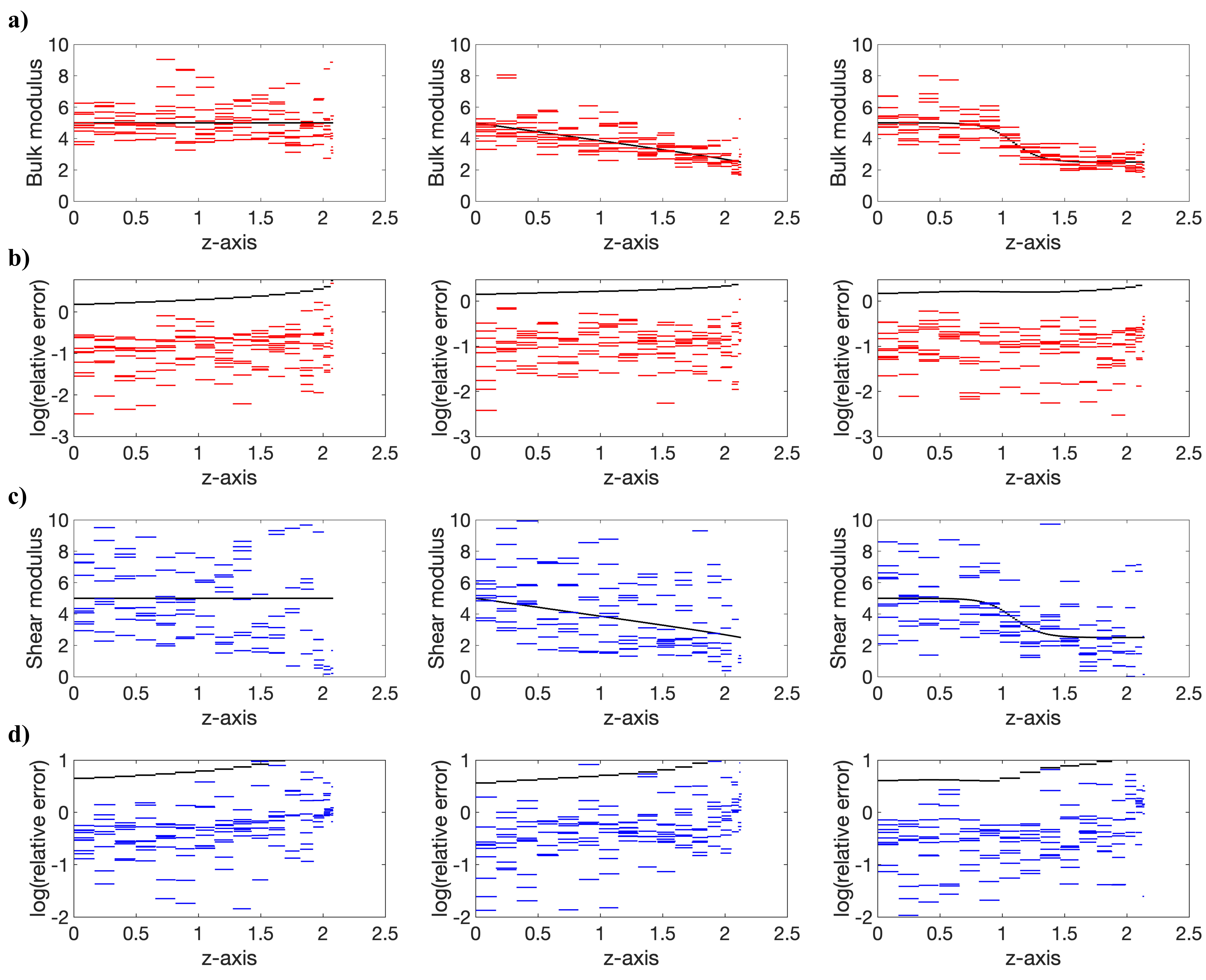}	
	\caption{\textbf{Inference of elastic moduli from noisy data.} Step-function inference from noisy data for bulk modulus $\textbf{a)}$ and shear modulus $\textbf{c)}$ using $N_I=16$ linear segments. Ten inferences based on different noisy data are shown, where the noisy data are generated by adding random perturbations of $1\%\times$ maximal cell radius to the synthetic cell wall profiles. 
		The relative errors in inferred bulk modulus \textbf{b)} and shear modulus \textbf{d)} are plotted in base-10-logarithmic scale, compared with the estimate of error bounds (black) based on the perturbation analysis. For better view, the rightmost error bound and some outliers of the inference are not shown for the shear modulus.
		Compared with inference results with $N_I=8$ segments (Fig.~\ref{fig:perturbedInference}), the inference with $N_I=16$ (finer discretization) has much larger relative errors and becomes unreliable.
		\label{fig:kmu_N16}}
\end{figure}

\begin{figure}[h]
	\centering
	\includegraphics[width=1\linewidth]{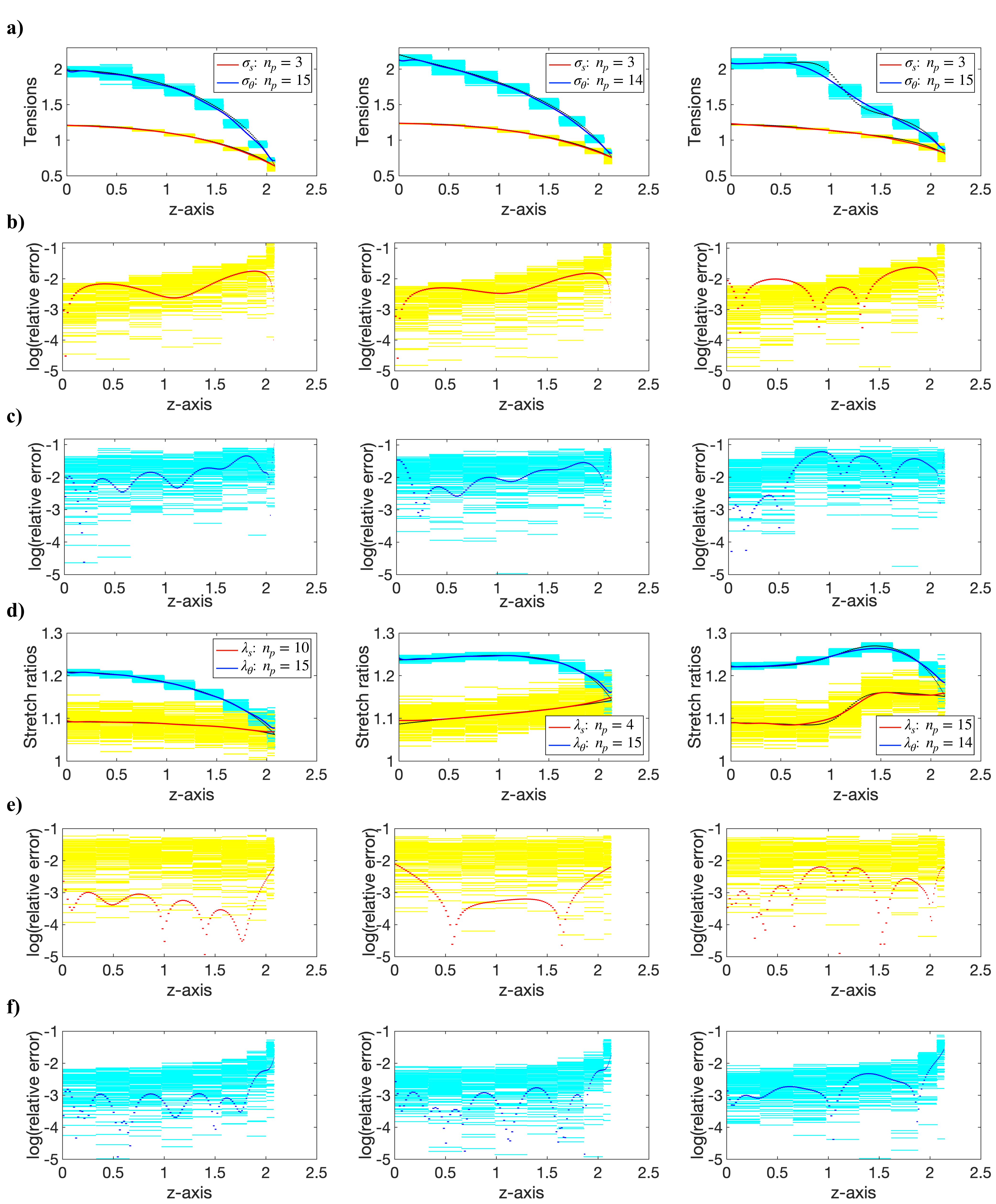}
	\caption{\textbf{Polynomial inference of canonical tensions and stretch ratios from multiple cell data.} 		
	The optimal polynomial inferences of the tensions $\textbf{a)}$ and stretch ratios $\textbf{d)}$ are obtained by the optimization scheme based on 100 cell outlines with $1\%$ random noise. On each cell outline, we use $N_I=8$ linear segments and the optimal degree of polynomial fitting $n_p$ is shown in the insets. The red and blue curves represent the polynomial inference of meridional and circumferential tensions or stretches; the yellow and cyan linear segments represent the step-function inference, respectively. 
	The relative errors in inferred tensions \textbf{b)} and \textbf{c)} and stretches \textbf{e)} and \textbf{f)} are plotted in $\log_{10}$ scale, where the colors are the same with $\textbf{a)}$ and $\textbf{d)}$. 
	\label{fig:optm_tensions}}
\end{figure}

\begin{figure}[h]
	\centering
	\includegraphics[width=1\linewidth]{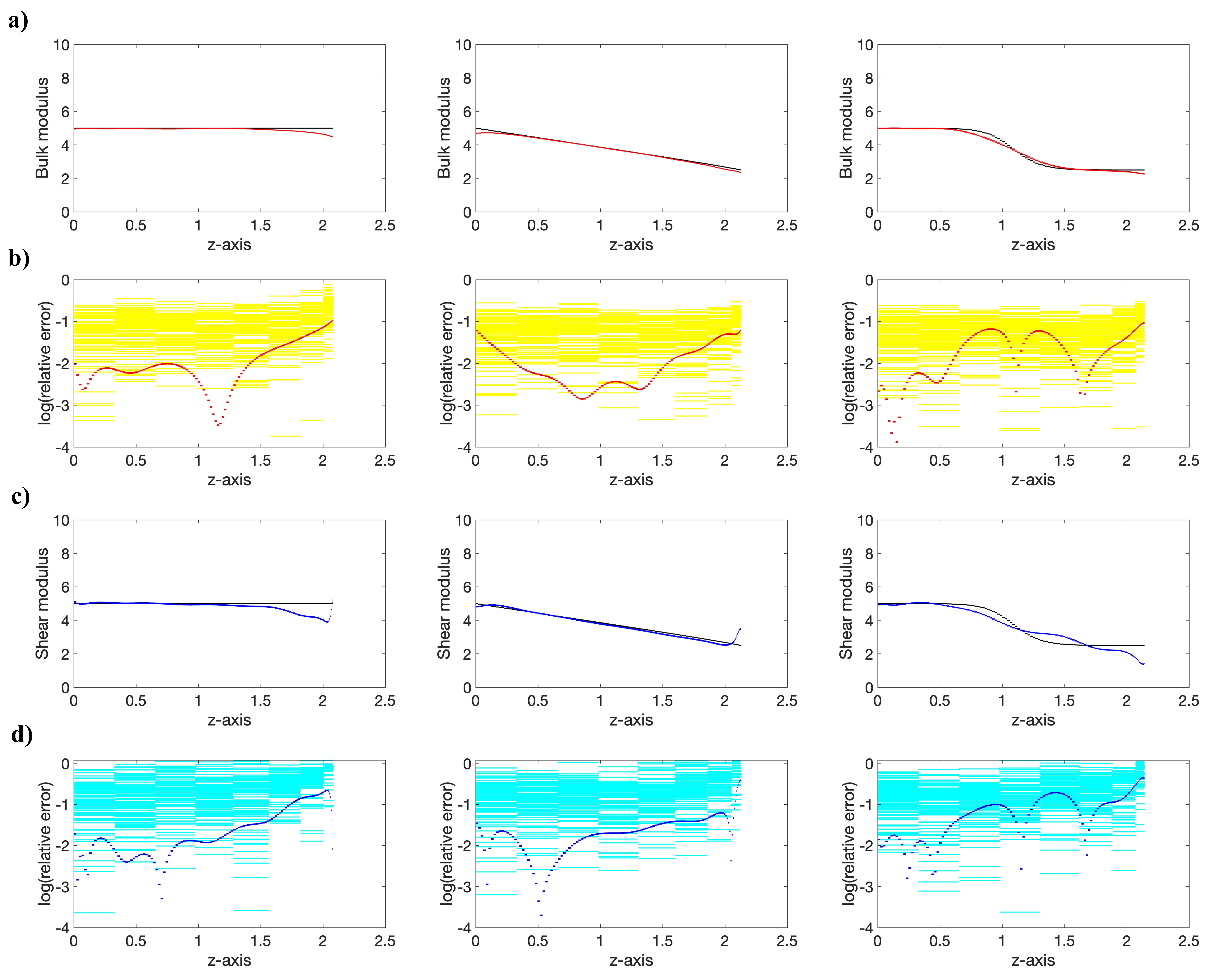}
	\caption{	\textbf{Inference of canonical elastic moduli by Approach 2.} 		
	The smooth-curve inferences of the bulk modulus $\textbf{a)}$ and shear modulus $\textbf{c)}$ are calculated based on the polynomial inferences of canonical tensions and stretches in Fig.~\ref{fig:optm_tensions}, known as Approach 2 of our optimization inference scheme (see main text). The inferred bulk modulus (red) and shear modulus (blue) are compared with the prescribed values (black). 
	The relative errors in inferred bulk modulus \textbf{b)} and shear modulus \textbf{d)} are plotted in $\log_{10}$ scale. The errors for smooth-curve inferences are in dotted curves and those for step-function inferences are in linear segments. 
	The accuracy of inference by Approach 2 does not significantly differ from that by Approach 1. 
	\label{fig:kmu_app2}}
\end{figure}
\newpage
\begin{figure}[h]
	\centering
	\includegraphics[width=1\linewidth]{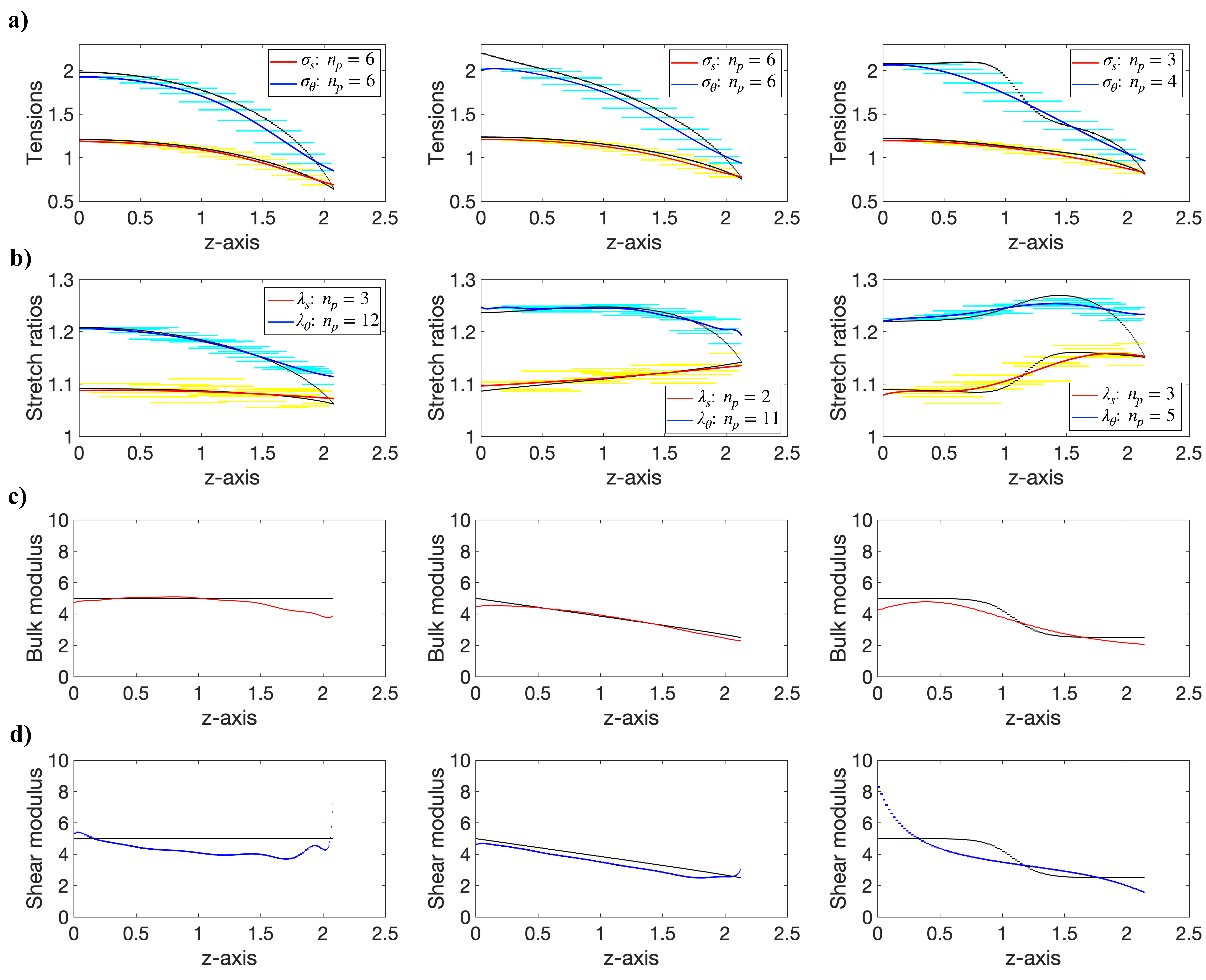}
	\caption{\textbf{The cost-effective inference on a single cell data with $1\%$ noise.}
	The optimal polynomial inferences of tensions \textbf{a)} and stretch ratios \textbf{b)} are obtained by fitting to the constant-value inferences, which are based on a set of 5 low-resolution data samples. See main text for details. The optimal degrees of polynomial are shown in the inset. The smooth-curve inferences for bulk modulus \textbf{c)} and shear modulus \textbf{d)} are calculated using the optimal polynomial inference of tensions and stretches via Eqs.(\ref{eq:continuous_K}) and (\ref{eq:continuous_mu}), and are shown with the prescribed values (black). 
	\label{fig:low_shifting}}
\end{figure}




\end{appendices}


\bibliography{ref}% common bib file


\end{document}